\documentclass[a4paper,11pt]{article}
\pdfoutput=1 % if your are submitting a pdflatex

\usepackage{jheppub} % for details on use of the package, please see the JHEP-author-manual

\usepackage[T1]{fontenc} % if needed
\usepackage[utf8]{inputenc}
\usepackage{cleveref,siunitx,hyperref,tikz,pgfplots}
\pgfplotsset{compat=newest}
\usepgfplotslibrary{external} % This makes external figures; delete the file to remake figure
\tikzsetexternalprefix{figures/}
\usetikzlibrary{decorations,decorations.text,arrows.meta,decorations.markings,
decorations.pathreplacing,fadings,decorations.pathmorphing,3d}
\pgfplotsset{legend image code/.code={\filldraw[#1] (-1mm,-1mm) rectangle (1mm,1mm);}, legend
cell align=left}

\renewcommand\d{\mathrm d}
\newcommand{\inv}[1]{\frac1{#1}}
\newcommand\cosec{\operatorname{cosec}}

\newcommand{\half}[1]{\frac{#1}2}

\title{Constraining active-sterile neutrino transition magnetic
moments at DUNE near and far detectors}

\author[a]{Thomas Schwetz,}
\author[a]{Albert Zhou,}
\author[a,b]{and Jing-Yu Zhu}
\affiliation[a]{Institut f\"ur Astroteilchen Physik, Karlsruher
Institut f\"ur Technologie (KIT),\\ Hermann-von-Helmholtz-Platz 1,
76344 Eggenstein-Leopoldshafen, Germany}
\affiliation[b]{School of Physics and Astronomy and Tsung-Dao Lee Institute,
Shanghai Jiao Tong University,\\ 800
Dongchuan Rd, Shanghai 200240, China}

\emailAdd{schwetz@kit.edu}
\emailAdd{albert.zhou@kit.edu}
\emailAdd{jing-yu.zhu@kit.edu}

\abstract{We consider the sensitivity of the DUNE experiment to a
heavy neutral lepton, HNL (also known as sterile neutrino) in the
mass range from a few MeV to a few GeV, interacting with the Standard
Model via a transition magnetic moment to the active neutrinos, the
so-called dipole portal. The HNL is produced via the up-scattering of
active neutrinos, and the subsequent decay inside the detector
provides a single-photon signal. We show that the tau-neutrino
dipole portal can be efficiently probed at the DUNE far detector,
using the tau-neutrino flux generated by neutrino oscillations,
while the near detector provides better sensitivity to the electron-
and muon-neutrino dipole portal. DUNE will be able to explore large
regions of currently unconstrained parameter space and has
comparable sensitivity to other planned dedicated experiments, such
as SHiP. We also comment briefly on the sensitivity to pure HNL
mixing with the tau neutrino at the DUNE far detector.}

\begin{document}
\maketitle
\flushbottom
%%%%%%%%%%%%%%%%%%%%%%%%%%%%%%%%%%%%%%%%%%%%%%%%%%%%%%%%%%%%%%%%%%%%%%%%%%%%%%%%
\section{Introduction} \label{sec:intro}%

Neutrino mass requires an extension of the Standard Model (SM). Many
known mechanisms to give mass to neutrinos involve fermionic
SM-gauge-group singlets, so-called sterile neutrinos, also referred to
as heavy neutral leptons (HNL). However, in general there are no specific
indications about the mass scale of such sterile neutrinos. Moreover,
it is possible that they act as a portal to generic new physics. In
this paper we will assume that a HNL, denoted by \(\nu_4\), exists in the mass
range from a few~MeV to a few~GeV, and that it interacts with the SM
via a transition magnetic moment --- the so-called ``dipole portal''
--- described by the following term in the Lagrangian:
\begin{equation}\label{eq:dipole}
	\mathcal L = d_\alpha \overline\nu_{\alpha L} \sigma^{\mu\nu} \nu_4 F_{\mu\nu} + {\rm h.c.}
\end{equation}
Here, $\nu_\alpha$ is a left-handed SM neutrino field of flavour
$\alpha = e,\mu,\tau$, $F_{\mu\nu}$ is the photon field strength
tensor, and $\sigma^{\mu\nu}=\frac{i}{2}(\gamma^\mu\gamma^\nu-
\gamma^\nu\gamma^\mu)$ is the antisymmetric combination of Dirac gamma
matrices.

\Cref{eq:dipole} corresponds to an effective Lagrangian, valid up to a
cut-off energy scale $\Lambda$, where the transition magnetic moment
$d_\alpha$ is expected to be of order $1/\Lambda$. Note that
\cref{eq:dipole} is not SU(2)$_L$ gauge invariant and therefore
$1/\Lambda \sim v/{\Lambda'}^2$, with $v$ denoting the Higgs vacuum
expectation value. A discussion on some model-building aspects can be
found in \cite{Magill:2018jla}, see also \cite{Babu:2020ivd}. In this
work we will remain agnostic about the UV origin of this operator and
study its phenomenological implications at energies small compared to
the electro-weak scale $v$.

The heavy-neutrino dipole portal has been investigated by a number of
authors. Comprehensive reviews of various laboratory, astrophysical,
and cosmological bounds on $d_\alpha$ can be found in refs.\ \cite{
  Magill:2018jla,Brdar:2020quo}. Other studies include considerations
of solar neutrinos \cite{Shoemaker:2018vii,Shoemaker:2020kji,
  Plestid:2020vqf}, atmospheric neutrinos in IceCube~\cite{
  Coloma:2017ppo}, or short-baseline
experiments~\cite{Gninenko:2009ks,
  Gninenko:2010pr,Masip:2012ke,Ballett:2016opr,Fischer:2019fbw,
  Vergani:2021tgc}. The HNL dipole portal can be explored also at
dedicated experiment for long-lived particle searches such as
SHiP~\cite{Anelli:2015pba},
FASER~\cite{Feng:2017uoz,Jodlowski:2020vhr}, or
MATHUSLA~\cite{Curtin:2018mvb,Alpigiani:2020tva}.

Typically it is difficult to test a transition moment between the tau
neutrino and a HNL, $\alpha=\tau$ in \cref{eq:dipole}, since it is
hard to produce an intense $\nu_\tau$ flux. In this work we will
exploit neutrino oscillations to overcome this problem: we consider
$\nu_\mu\to\nu_\tau$ oscillations at the DUNE long-baseline experiment
\cite{Abi:2020evt}, governed by a transition amplitude of order
one. These $\nu_\tau$ may up-scatter on nuclei, nucleons or electrons to a HNL
via the dipole interaction \cref{eq:dipole}. The heavy neutrino can
travel over macroscopic distances and decay back into a light neutrino
and a photon inside the detector. Below we will calculate the
sensitivity of the DUNE far detector to $d_\tau$ using these
processes. Due to the sizeable primary $\nu_\mu$ and $\nu_e$ fluxes,
the HNL transition moments $d_\mu$ and $d_e$ are more efficiently
probed at the near detector. We will also provide estimates of the
DUNE near detector sensitivities from the $\nu_{\mu,e}$-up-scattering
processes.

The outline of our paper is as follows. In \cref{sec:signal} we
discuss the general features of the signal considered in this paper
and provide an outline of the relevant event-rate calculations.
\Cref{sec:results} contains our main results: in \cref{subsec:far-det}
we show the sensitivity of the DUNE far detector to the tau-neutrino
dipole portal, whilst \cref{subsec:near-det} contains the
near-detector sensitivities to the electron- and muon-neutrino dipole
portal. In \cref{subsec:global} we set the DUNE sensitivities in the
context of various laboratory and astrophysical constraints from the
literature, showing that DUNE will cover large currently unconstrained
regions in parameter space, and is competitive with prospective
sensitivities from the SHiP experiment \cite{Anelli:2015pba}. In most
parts of this paper we will assume that HNL mass mixing with active
flavours is negligible and the dipole interaction dominates. However
in \cref{sec:mixing} we briefly comment on the sensitivity of the DUNE
far detector to HNL mixing with $\nu_\tau$: we find that the far
detector sensitivity is somewhat weaker than the sensitivity of the
near detector from the prompt $\nu_\tau$ flux. We conclude in
\cref{sec:conclusion}. In \cref{app:xsec} we summarize the cross
section formulae relevant for the HNL up-scattering mediated by the
dipole portal. \Cref{app:ins-ev-intgrl,app:out-ev-intgrl} provide
technical details on the event-rate calculations. In
\cref{app:dipole-ND} we estimate the event rate at the near detector
from the tau-neutrino dipole portal, indicating that the signal at the
far detector dominates.

Throughout this article, the HNL decay width we use applies in the case
of a Dirac sterile neutrino. A Majorana neutrino will have twice as
large a decay width \cite{Balantekin:2018ukw}. A treatment of their
interesting differences can be found in \cite{Balantekin:2018azf,
Balantekin:2018ukw,Berryman:2019dme}.

%%%%%%%%%%%%%%%%%%%%%%%%%%%%%%%%%%%%%%%%%%%%%%%%%%%%%%%%%%%%%%%%%%%%%%%%%%%%%%%%
\section{Dipole decay signal at DUNE}\label{sec:signal}%

\begin{figure}[t]
	\centering
	\includegraphics{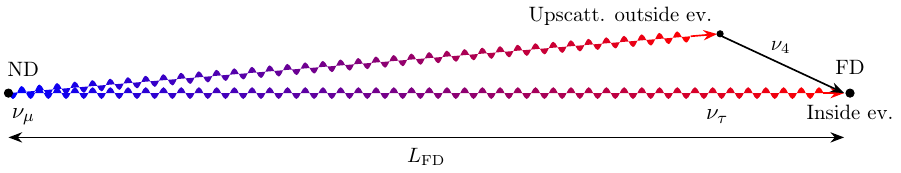}
	\includegraphics[width=.75\textwidth]{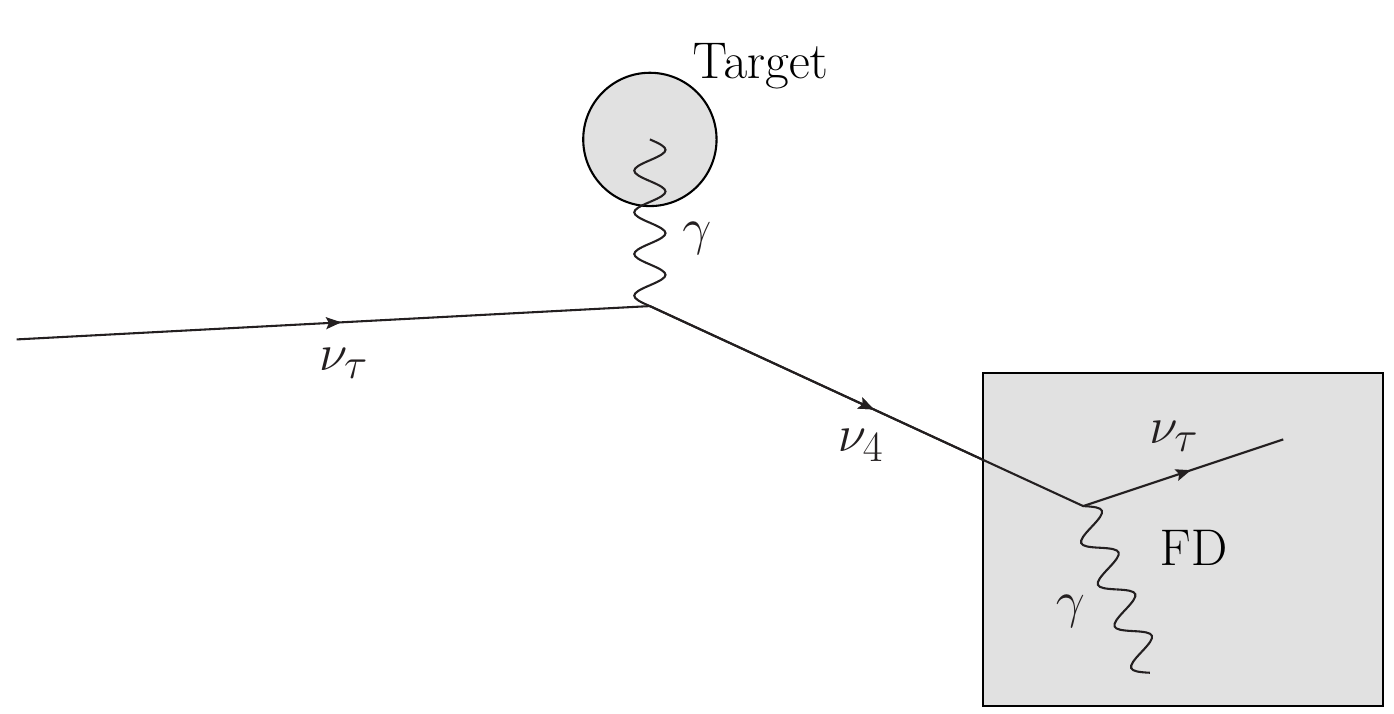}
	\caption{Cartoons of the up-scattering production and decay signal for $d_\tau$ in the far
	detector. \label{fig:signal}}
\end{figure}

The DUNE experiment produces a flux of mostly muon neutrinos with
a subleading component of electron neutrinos. These fluxes can be
used to directly search for $d_e$ and $d_\mu$ at the DUNE near
detector (ND). However, to constrain a tau transition moment $d_\tau$,
we use the muon neutrinos, which oscillate into tau neutrinos during
their propagation through the Earth.\footnote{The primary flux of tau
neutrinos in the beam has been estimated recently in
\cite{Berryman:2019dme,Coloma:2020lgy,Breitbach:2021gvv}. We have
estimated that the sensitivity of the $d_\tau$-induced event rate in
the ND is significantly smaller than the one in the FD discussed here,
see \cref{subsec:near-det,app:dipole-ND} for further discussions.} Tau
neutrinos can up-scatter off target particles in the Earth (protons,
electrons or coherently on nuclei) and can be converted via the dipole
transition moment into a sterile neutrino. If the up-scattering occurs
outside the detector in the Earth's crust or upper mantle, and if the
sterile-neutrino mass is low enough, the sterile neutrino will be
long lived and can travel through the Earth and decay inside the DUNE far detector (FD).
We call these \emph{outside} events. If the sterile neutrino has a
large enough mass and/or dipole moment, both up-scattering and decay
can occur inside the detector. We call these \emph{inside} events. The
sequence of events is illustrated in \cref{fig:signal}.

We consider coherent scattering off nuclei in the Earth and incoherent
scattering off protons, neutrons and electrons. Note that, since
neutrons can only interact magnetically, as they have no electric
charge, their cross section is suppressed by a factor \(\eta\equiv Q^2/
(4M_T^2)\). Our expressions come from \cite[App. A]{Magill:2018jla}
and \cite[Eqs. (4,5)]{Brdar:2020quo}, and are summarised for
completeness in \cref{app:xsec}. For neutral-current (NC) scattering,
the cross-section for electrons can be neglected, since the total
cross section scales linearly with the mass of the target particle,
see also \cite[Fig.\ 8]{Tomalak:2019ibg}. However, the total dipole
cross section goes logarithmically with the target-particle mass, so
electrons can give a relevant contribution (see below). For the
coherent cross section, we consider Argon nuclei for the events inside
the detector, and for the outside events the nuclei: oxygen, silicon,
aluminium, iron, calcium, sodium, magnesium, potassium, titanium;
whose abundances we take from \cite{Haynes}.

Therefore, depending on the different event classes, we expect the
following signatures in the detector:
\begin{itemize}
	\item {\bf Outside events:} As the scattering happens outside the
		detector, the signature is a single-photon event.
	\item {\bf Inside events, coherent:} The coherent scattering on the
		nucleus leaves the nucleus intact and provides a nuclear recoil of
		low energy, which is difficult to observe in the detector (see however
		\cite{Sangiorgio:2013ib,Agnes:2018mvl}). Thus, the signal is again
		a single-photon event.
	\item {\bf Inside events, incoherent:} The incoherent scattering on
		nucleons leads to a signature similar to NC neutrino events, whereas
		the scattering on electrons results in a single electron. Hence, the
		signature will be either a NC-like or single-electron type event in
		coincidence with the displaced single-photon event from the
		heavy-neutrino decay.
\end{itemize}%

While the first two event classes provide a clean single-photon
signal, the last event category offers a richer signature which can be
explored by dedicated analyses for improved background discrimination.
In this work we will consider all event types on equal footing and
assume for all of them the single-photon reconstruction efficiency
given in \cite{Abi:2020evt} (see \cref{app:ins-ev-intgrl} for further
details). Liquid-argon detectors can discriminate electrons and photons
very well \cite{Rubbia:2011ft,Acciarri:2015ncl,Adams:2018sgn}; so the
relevant backgrounds for the dipole signal are the single-photon
process NC1\(\gamma\) and highly asymmetric NC\(\pi^0\)-decays, where
the two photons from the pion decay cannot be distinguished.
Single-photon production from neutral-current scattering
(NC1\(\gamma\)) is expected to have a total cross section
\(\sim\SI{e-41}{cm^2/nucleon}\) \cite{Garvey:2014exa, Zhang:2012xn,
Wang:2013wva}. We estimate that this process would induce a background
of \SI{0.1}{events/year} (see \cite{Abe:2019cer} for a NC1\(\gamma\)
analysis in T2K). Therefore, photons from NC$\pi^0$ are expected to be
the dominant background, see \cite{Adams:2018sgn,Acciarri:2015ncl} for
studies in liquid argon detectors. A detailed detector simulation
including analysis cuts is beyond the scope of this work and we
concentrate on predicting the expected signal rate.

Below we outline the calculation of the event rate. We focus first on
the $d_\tau$-induced signal in the FD and comment on the relevant
modification for the $d_{\mu,e}$ signal in the ND in
\cref{subsec:near-det}. We use the indices \(\nu\), 4, \(T\), and
\(\gamma\) to denote the initial light neutrino, the heavy sterile
neutrino, the scattering target, and the final photon, respectively.
The target $T$ can be an electron, proton, neutron, or a nucleus (for
coherent interactions).

%%%%%%%%%%%%%%%%%%%%%%%%%%%%%%%%%%%%%%%%%%%%%%%%%%%%%%%%%%%%%%%%%%%%%%
\subsection{Inside-event rate}

Let a sterile neutrino be produced at the location \(\vec x_p\) and
decay at \(\vec x_d\), both inside the detector volume $V_D$; let
\(\ell_d\) be the maximum decay length, when \(\vec x_d\) is at the
edge of the detector. The decay direction (which defines the solid
angle \(\Omega_s\)) is \((\vec x_d-\vec x_p)/\ell_d\). We wish to
calculate the following integral for the differential event rate from a
given target type $T$:
\begin{equation} \label{eq:ev-rate}
	\frac{\d N_T}{\d E_\nu}=N_\text{mod}\int_{V_D}\hspace{-2ex}\d^3\vec{x}_p\frac{L_\text{ND}^2}
	{|\vec{x}_p|^2}\frac{\d^2\Phi}{\d\Omega\d E_\nu}P_\text{osc}\left(\frac{|\vec x_p|}{E_\nu}
	\right)\rho_N(\vec{x}_p)\int\d\Omega_s\frac{\d\sigma_T}{\d\Omega_s}P_\text{dec}(\ell_d)
	\varepsilon(p_4) \,.
\end{equation}
From left to right, the terms on the RHS are the number of detector
modules \(N_\text{mod}=4\), an integral of the up-scattering location,
the geometric suppression, the differential $\nu_\mu$ flux $\Phi$ at
the near-detector location $L_{\rm ND}$, the $\nu_\mu\to\nu_\tau$
oscillation probability $P_{\rm osc}$, the nucleon density $\rho_N$,
the integral over the scattering solid angle, the cross section per
solid angle per \emph{nucleon}, the probability $P_{\rm dec}$ that the
heavy neutrino decays inside the detector, and the photon
reconstruction efficiency $\varepsilon$ evaluated as a function of the
heavy-neutrino momentum. The energy of the light neutrino is $E_\nu$,
whereas \(E_4\) and \(p_4\) denote the energy and momentum of the
heavy neutrino. Details of all these terms may be found in
\cref{app:xsec,app:ins-ev-intgrl}.

We make a series of approximations, which are also detailed in
\cref{app:ins-ev-intgrl}. Briefly, as the baseline is much larger than
the detector dimensions we set \(|\vec x_p|=L_\text{FD}\) and assume
the neutrino beam is collimated (\emph{i.e.\ }on-axis), so there is no
angular dependence of the flux. We also take the detector nucleon
density \(\rho_N\) to be constant (whose value is determined by the
detector density).

By assuming the detector is cylindrical and using cylindrical
coordinates \((\rho,\theta_p,z) \), the polar-angle dependence,
\(\theta_p\), of the integrand drops out. By assuming the decay length
\(\ell_d\) can be approximated by its value at the centre
\(\ell_d(\rho=0)\equiv \ell_d^0\), the \(\rho\)- and
\(\varphi_s\)-integrals can be done analytically. This approximation
becomes exact in the limit of large decay width (as all decays happen
instantaneously so \(\ell_d\) becomes irrelevant), but when the
decay-length is comparable to or larger than the detector size there
should be a penalty term \(\Pi\left(\ell_d^0\right)\) due to the
geometry; this is precalculated (see \cref{app:ins-ev-intgrl}). With
this simplifying assumption, the \(\rho\)-\(\theta_p\) integral
\(\iint_A\rho\d\rho\d \theta_p\) may simply be replaced with the
detector cross-section area \(A_\text{det}\). The final expression,
summed over the scattering targets $T$, is then
\begin{equation}\label{eq:spect-inside}
	\begin{aligned}
		\frac{\d N}{\d E_\nu}=&N_\text{mod}\frac{L_\text{ND}^2}{L_\text
		{FD}^2}\rho_NA_\text{det}\left.\frac{\d\Phi}{\d\Omega\d E_\nu}
		\right|_{\theta_b=0}P_\text{osc}\left(\frac{L_\text{FD}}{E_\nu}
		\right)\times \\
		& \sum_{M_T}\int_0^{L_d}\d z \int_{-1}^1\d\cos\theta_s\frac{\d
		\sigma_T}{\d\cos\theta_s}\Pi\left(\ell_d^0\right)P_\text{dec}
		(\ell_d^0)\varepsilon(p_4)\,.
	\end{aligned}
\end{equation}%

%%%%%%%%%%%%%%%%%%%%%%%%%%%%%%%%%%%%%%%%%%%%%%%%%%%%%%%%%%%%%%%%%%%%%%%%%%%
\subsection{Outside-event rate}%

We start from \cref{eq:ev-rate}, which also applies to the outside events; however we will now
integrate \(\vec x_p\), the \(\nu_4\) production point, over the region of the Earth exposed
to the neutrino beam including the off-axis beam \cite{Abi:2020wmh,DUNE-off-axis-flux}. In
addition, the nucleon density is assumed to be constant below sea level and zero above sea
level. Inside the Earth, we take \(\SI{2.9}{g/cm^3}\) as the mass density, which is roughly
the surface density (although at \(\theta_b^\text{max}\approx3.6^\circ\) the density can reach
\SI{3.375}{g/cm^3} \cite{Dziewonski:1981xy}).%

Note that the geometry of the Earth breaks the otherwise cylindrical symmetry around the
beam axis. We denote the coordinates of the scattering point in spherical coordinates \((r_p,
\theta_b, \varphi_b)\); then the boundary of the earth imposes the boundary conditions on the
\(\varphi_b\)-integral: \(\varphi_b\in\left[\Phi(r_p,\theta_b),2\pi-\Phi(r_p,\theta_b)\right]
\) (see \cref{fig:outside-3d-pic} in \cref{app:out-ev-intgrl}). Since the rest of the
integrand has no \(\varphi_b\)-dependence, we replace the \(\varphi_b\)-integral with the
factor \(\varepsilon_{\varphi_b} \equiv2\left[\pi-\Phi(r_p,\theta_b)\right]\). Finally, we
assume the geometry of the detector can be neglected. Therefore, \(\theta_s\) and
\(\varphi_s\) are uniquely determined by the production point \(\vec x_p\). We replace the
inner integral with an estimate of the solid angle \(\Delta\Omega_s\) of the detector as
viewed from \(\vec x_p\).
Note that the decay probability also changes to account for the distance required to travel to
the detector. For expediency, when the phase of the
oscillation probability is large, we replace \(P_\text{osc}\) with \(1/2\), assuming the
wiggles average out; this was checked to only negligibly change the result compared to the
full integral.
%Furthermore, we evaluate the spectrum in terms of \(E_4\).
%The reconstruction efficiency is the same as the inside-case.
The final expression is
\begin{equation}\label{eq:spect-outside}
	\begin{aligned}
		\frac{\d N}{\d E_4} = &N_\text{mod}\frac{\rho_N}{2\pi}\int_0^
		{\theta_b^\text{max}}\sin\theta_b\d\theta_b\int_{r_\text{min}}^
		{r_\text{max}}L_\text{ND}^2\d r_p\times \\
		&\sum_{M_T}\left[\frac{\d^2\Phi}{\d\Omega_b\d E_\nu}\frac{\d
		E_\nu}{\d E_4}P_\text{osc}\left(\frac{r_p}{E_\nu}\right)\varepsilon_
		{\varphi_b}\left(r_p,\theta_b\right)P_\text{decay}(\ell)\frac{\d
		\sigma}{\d\cos\theta_s}\Delta\Omega_s\varepsilon(p_4)\right]_T.
	\end{aligned}
\end{equation}
The evaluation of \(\varepsilon_{\varphi_b}\), as well as more details on the off-axis flux,
geometric relations and approximations can be found in \cref{app:out-ev-intgrl}; kinematic
relations can be found in \cref{app:xsec}. We use the same reconstruction efficiency given in
\cref{app:ins-ev-intgrl}.%

%%%%%%%%%%%%%%%%%%%%%%%%%%%%%%%%%%%%%%%%%%%%%%%%%%%%%%%%%%%%%%%%%%%%%%%%%%%%%%%%%%%%%%%
\subsection{Example spectra}%

\begin{figure}[t]
	\includegraphics{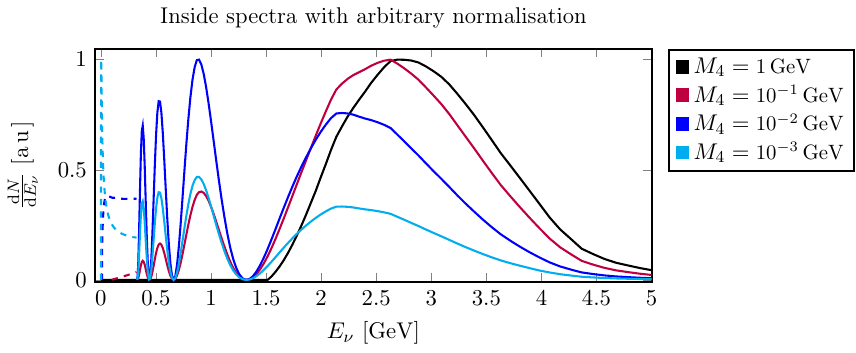}
	\includegraphics{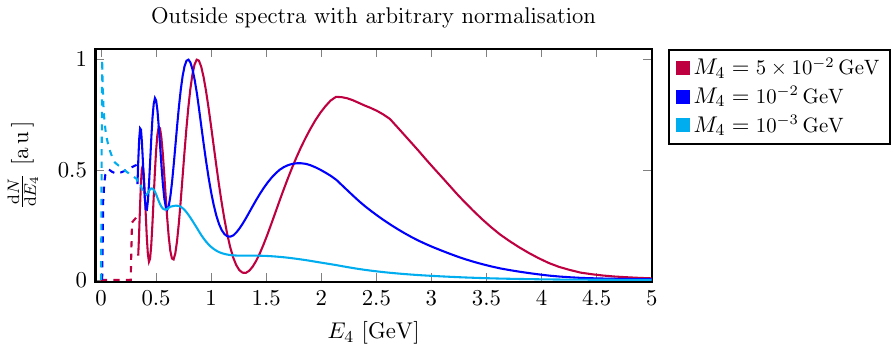}
	\caption{Spectra for inside and outside events at various
					masses, normalised so that the peak is 1. At low energies
					(dashed), we replace the oscillation probability with
					\(1/2\) to account for the averaging of fast oscillations.}
	\label{fig:spectra}
\end{figure}%

In \cref{fig:spectra}, we show example spectra for heavy-neutrino
masses \(M_4=10^{-3},10^{-2},10^{-1},1~\si{GeV}\) for inside events
and \(M_4=10^{-3},10^{-2},5\times10^{-2}~\si{GeV}\) for outside
events. For low energies (dashed), we replace the oscillation
probability with \(1/2\) to account for the averaging of fast
oscillations. For the inside events this is only done for the plots
shown and is not used for the numerical analysis, whereas for the
outside events this significantly speeds up calculations and has a
negligible impact on the final result.

We show inside events as a function of light-neutrino energy $E_\nu$,
whereas the outside events as a function of the heavy-neutrino energy
$E_4$. This is due to calculational convenience.\footnote{For inside
events, the flux and oscillation probability (as functions of $E_\nu$)
can be factored out of the inner integrals, whereas for the outside
events, we use the more convenient expression of \(E_\nu(E_N)\) see
\cref{eq:spect-inside,eq:spect-outside}.} The total number of events
is then obtained by integrating over these spectra. We remark that for
the lowest masses \(M_4\sim\SI1{MeV}\), there is a sharp peak at low
energies. This is due to the \(1/Q^2\) term in the cross section. The
DUNE reconstruction efficiency for photons remains high at low photon
momentum (0.7 at \(p_N=\SI{0.1}{GeV}\)~\cite{Abi:2020evt}). For this
reason, we do not cut off the spectrum at low energies and instead
just fold in the reconstruction efficiency, see
\cref{app:ins-ev-intgrl}.%

%%%%%%%%%%%%%%%%%%%%%%%%%%%%%%%%%%%%%%%%%%%%%%%%%%%%%%%%%%
\section{Results}\label{sec:results}%

\subsection{DUNE-FD sensitivity to \texorpdfstring{$d_\tau$}{the tau
transition moment}} \label{subsec:far-det}%

\begin{figure}[t]
	\includegraphics{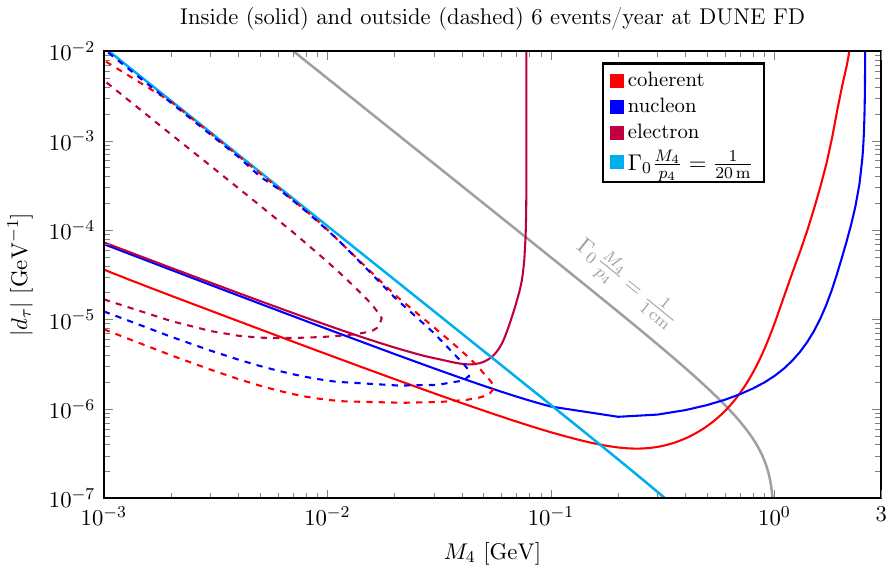}
	\caption{The 6-events/year curve for inside (solid) and
					outside (dashed) events at the DUNE FD for coherent
					scattering on nuclei (red), incoherent scattering
					on nucleons (blue) as well as electrons (purple). Near the
					upper curves of the outside events, up-scattering occurs
					close to the detector and our approximations break down, as
					indicated by the cyan line, at which the decay length is
					\SI{20}m for \(E_4=\SI1{GeV}\). This effect is negligible
					as inside-events will dominate. We also show in grey when
					the decay-length is of the order of DUNE's spatial
					resolution \cite{Abi:2020loh}.}
	\label{fig:DUNE-FD-events}
\end{figure}%

In \cref{fig:DUNE-FD-events} we show the six-events/year curve for
inside (solid) and outside (dashed) events at the DUNE far detector
for coherent scattering on nuclei (red), incoherent scattering
on nucleons (blue) as well as electrons (purple) as a function of the
heavy-neutrino mass $M_4$ and transition magnetic moment $d_\tau$. Due
to the lightness of the electron, up-scattering is kinematically
allowed for low masses only, hence why the corresponding curve for
inside events cuts off at \(M_4\sim\SI{0.077}{GeV} \). The coherent
scattering dominates until \(M_4\sim\SI{0.7}{GeV}\), after which
incoherent scattering off nucleons dominates. However above \SI1{GeV}
the available beam energy rapidly kills the number of events.

For the outside events, the upper parts of the curves are indicative
only. As \(d_\tau\) increases, the decay length \(1/\Gamma\)
decreases, so that more events occur close to the detector. At some
point the events become too clustered around the detector, such that
either the integrator cannot resolve the sharp peak
and/or our assumptions break down (we begin to resolve the detector
geometry). The cyan curve in the figure indicates when the decay
length is \SI{20}m at \(E_4=\SI1{GeV}\);
for \(d_\tau\) roughly above this line a full treatment of the detector
geometry would be needed. Since inside events will dominate in this region
anyway, we safely ignore this point.

DUNE will have sub-centimetre spatial resolution
\cite{Abi:2020loh}. The grey curve in \cref{fig:DUNE-FD-events}
corresponds to the case where the decay length for a HNL with energy
1~GeV is 1~cm. Hence, in the region roughly below the grey
curve it will be possible to resolve the displaced vertices of the
hadronic signal and the single-photon decay for inside events. Note
that the cyan and grey curves are only indicative, since they are for
fixed HNL energy of $E_4 = 1$~GeV and do not take into account the
energy spectrum.

\begin{figure}[t]
	\includegraphics{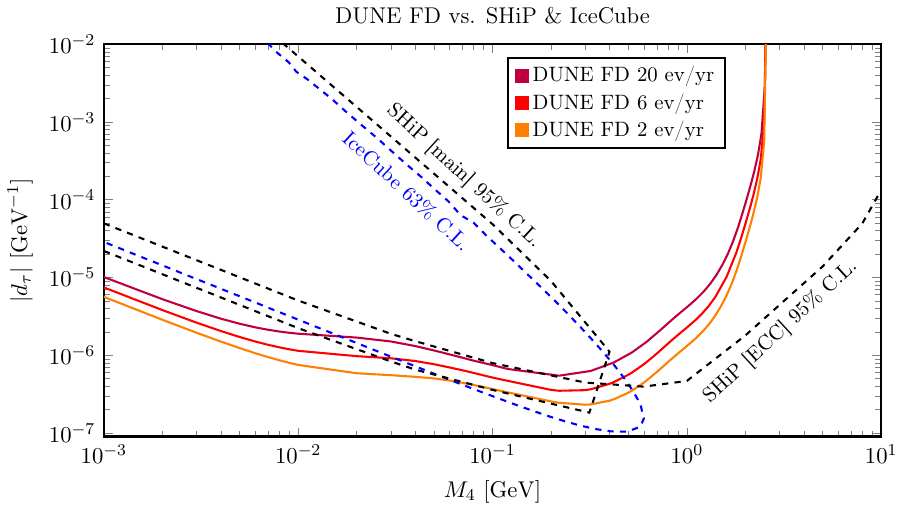}
%	\tikzsetnextfilename{SHiP-compare}
%	\begin{tikzpicture}
%		\begin{loglogaxis}[title={DUNE FD vs. SHiP \& IceCube}, xlabel=
%			{\(M_4~[\si{GeV}]\)},ylabel= {\(|d_\tau|~[\si{GeV^{-1}}]\)},
%			style={line width=1pt},legend style={{at=(.8,.85)},anchor=east},
%			unit vector ratio=2.5 1,unit rescale keep size=false,
%			ymax=1e-2,ymin=.9e-7, xmin=1e-3, xmax=10,width=.975\textwidth]
%			\addplot [purple] table [col sep=comma]{data/dune-curve-in-out-20ev.csv};
%			\addplot [red] table [col sep=comma]{data/dune-curve-in-out-6ev.csv};
%			\addplot [orange] table [col sep=comma]{data/dune-curve-in-out-2ev.csv};
%			\addplot [blue,dashed] table [col sep=comma]{data/dipole-icecube.csv};
%			\addplot[draw=none, postaction={decorate,decoration={text along path, text={{\small
%			IceCube 63\% C.L.}{}},text color=blue,raise=1ex, text align={center}}},domain=1e-2:1e-1]
%			{10^(-7.5)/x^(2.3)};
%			\addplot[draw=none, postaction={decorate,decoration={text along path, text={{\small SHiP
%			[main] 95\% C.L.}{}}, raise=1ex, text align={center}}}, domain=2e-2:2e-1]
%			{10^(-6.6)/x^(2.3)};
%			\addplot[draw=none, postaction={decorate,decoration={text along path, text={{\small SHiP
%			[ECC] 95\% C.L.}{}},text align={center}}}, domain=1:10, ] {10^(-6.9)*x^(2.3)};
%			\addplot [black,dashed] table [col sep=comma]{data/SHiP-curve-main-det.csv};
%			\addplot [black,dashed] table [col sep=comma]{data/SHiP-curve-ECC-det.csv};
%			\legend{\small DUNE FD 20 ev/yr, \small DUNE FD 6 ev/yr, \small DUNE FD 2 ev/yr}
%		\end{loglogaxis}
%	\end{tikzpicture}
	\caption{2-, 6-, 20-events/yr curves at the DUNE far detector
	(orange, red, purple, respectively). For comparison we show the
	95\%~C.L.\ sensitivities at SHiP from \cite{Magill:2018jla} assuming
	100 background events over 5 years of exposure, equivalently
	\SI{2e20}{POT}, (dashed black) for two detector types (ECC and main
	detector). In dashed blue is the one double-bang event/6~yrs curve
	at IceCube from \cite{Coloma:2017ppo}.}
	\label{fig:SHiP-compare}
\end{figure}%

Finally, we sum up the contributions from all event classes and show in
\cref{fig:SHiP-compare} the curves corresponding to 2, 6, and 20
signal events per year. We compare these curves to the
95\%~confidence-level sensitivities from the SHiP experiment. SHiP
\cite{Anelli:2015pba} is a proposed experiment at the Super Proton
Synchrotron (SPS) at CERN. Its sensitivity to the tau-neutrino's
dipole transition moment was evaluated in \cite{Magill:2018jla}.
These authors consider two type of detectors, an emulsion cloud
chamber (ECC) close to the beam target, as well as the ``main''
detector and calculate the 95\%~C.L.\ sensitivity after 5 years of
running (\SI{2e20}{POT} \cite{SHiP:2020hyy}), assuming 100 background
events in both detectors. For comparison, the curves shown for the
DUNE FD correspond to 10, 30 and 100 events in 5 years; using
signal/$\sqrt{\text{background}} \approx 2$ this would give a
95\%~C.L.\ sensitivity for 25, 225 and 2500 background events,
respectively.

In \cref{fig:SHiP-compare} we also compare our DUNE sensitivity with
an exclusion estimate for the IceCube detector due to so-called
double-bang events, induced by heavy neutrinos produced by atmospheric neutrinos
and decaying via the dipole operator \cite{Coloma:2017ppo}. The authors
calculate the curve for 1~event/6 years, corresponding to an exclusion
of \(1-e^{-1}\approx 63\%\) assuming no background. We note that the
original curve used \(\mu_\text{tr}\equiv 2d\) and units of
\(\mu_B\approx\SI{296.1}{GeV^{-1}}\).%

%%%%%%%%%%%%%%%%%%%%%%%%%%%%%%%%%%%%%%%%%%%%%%%%%%%%%%%%%%%%%%%%%%%%%%%%%%%%%%%%
\subsection{Near-detector sensitivity to the \texorpdfstring{$\nu_e$}{electron neutrino} and
\texorpdfstring{$\nu_\mu$}{muon neutrino} dipole portal} \label{subsec:near-det}%

It is possible to use the intense muon-, and somewhat-less-intense
electron-neutrino fluxes to constrain $d_e$ and $d_\mu$ at the near
detector. We consider a signal from up-scattering to the HNL in close
analogy to the FD signal described in \cref{sec:signal}. Our method
used earlier to calculate the event rates at the far detector (for
both inside and outside events) can be applied to this situation with
the following modifications: the baseline becomes \(L_\text{ND} =
\SI{574}m\), the oscillation probability is set to one, and we use the
intrinsic muon- and electron-neutrino fluxes from \cite{
DUNE-off-axis-flux}. We consider only the ND on-axis configuration and
assume the detector fiducial volume is \SI6m wide, \SI2m high and
\SI3m long with a fiducial mass of \SI{50}t \cite[§2.7]{
AbedAbud:2021hpb}. The crust at the surface of the Earth has a
density of \SI{2.6}{g/cm^3} \cite{Dziewonski:1981xy}, whereas soil has
a density \numrange{1.1}{1.6} \si{g/cm^3} \cite{soil-book}; we take an
average density of \SI2{g/cm^3} and we exclude the region inside and
next to the decay pipe \(r_p\geq \SI{270}m\) (see \cite[Fig.\ 1.2]{
Abi:2020wmh}); but we still assume a point-like neutrino source.
(This is conservative, since decays along the pipe will only lessen
the \(1/r_p^2\) geometric suppression.) The near detector has a depth
of \SI{62}m, so the angle of the beam with respect to ground level is
\(\operatorname{arcsine}(\SI{62}m/\SI{574}m)\approx\SI{108}{mrad}>
\theta_b^\text{max}\equiv\SI{62.72}{mrad}\), so we can safely set
\(\varepsilon_{\varphi_b}\) to one.

Our results for the $d_e$ and $d_\mu$ transition moments are shown
below in \cref{fig:global-pic-e,fig:global-pic-mu}, respectively, in
comparison with current limits as well as future sensitivities (see
\cref{subsec:global} for further discussion). We find that the DUNE-ND
six-events/year curves cover unconstrained parameter space in both cases
and are competitive with various prospective sensitivities.

Some comments are in order. Our analysis takes into account only HNL
production via the up-scattering process. As discussed in
\cite{Magill:2018jla} there will also be direct production from meson
decays produced in the target, either mediated by a virtual neutrino
or photon: $P^\pm\to\ell^\pm\overset{\scriptscriptstyle(-)}{\nu}
\hspace{-3pt}_4\gamma$, $P^0\to\gamma\nu_\alpha\nu_4$, where $P$
denotes a pseudo-scalar meson and $\ell$ a charged lepton. If the HNL
is long-lived enough to reach the detector, such processes will give
an additional contribution to the signal. The calculation of these
events would involve a detailed consideration of meson production in
the beam target, which is beyond the scope of this work. Note that
these processes will give only an additional contribution to the
signal; therefore our sensitivities are conservative in this respect.
However, due to the large $\nu_\mu$ flux at the ND, one expects also
significantly more SM NC events than in the FD, leading to a larger
background from NC$\pi^0$ events.

In principle one may expect also some prompt flux of tau neutrinos
produced at the beam target from heavy-meson decays. This could be
used, for instance, in the proposed SHiP experiment to search for HNLs
mixed with $\nu_\tau$ \cite{Anelli:2015pba,SHiP:2018xqw} and has been
considered in \cite{Magill:2018jla} to calculate the SHiP sensitivity
to the dipole portal $d_\tau$. At DUNE the beam energy is lower than
at SHiP and one expects a much reduced $\nu_\tau$ flux. Indeed, there
is no tau-neutrino flux available in the flux files provided by DUNE
\cite{DUNE-off-axis-flux}. The $\nu_\tau$ flux generated at the DUNE
beam has been estimated recently in \cite{Berryman:2019dme,
Coloma:2020lgy,Breitbach:2021gvv}. In principle this flux could be
used also to constrain $d_\tau$ at the DUNE ND. Extrapolating the
results from \cite{Coloma:2020lgy} we have estimated, however, that
the signal from the direct $\nu_\tau$ flux in the near detector would
be subleading to the one due to $\nu_\mu\to\nu_\tau$ oscillations in
the far detector considered in the present work (see
\cref{app:dipole-ND}). Therefore we concentrate here on the FD for
$d_\tau$.

On the other hand, the $\nu_\mu$ and $\nu_e$ fluxes (after taking into
account oscillations) will also produce a HNL signal from $d_\mu$ and
$d_e$ in the FD; this signal for $d_\mu$ should be comparable in
strength to the one for $d_\tau$ and somewhat weaker for
$d_e$. Therefore, comparing the ND sensitivity curves from
\cref{fig:global-pic-e,fig:global-pic-mu} to the one from the FD in
\cref{fig:global-pic-tau}, we expect the ND to provide superior
sensitivity to $d_e$ and $d_\mu$, due to much larger fluxes at the ND.

%%%%%%%%%%%%%%%%%%%%%%%%%%%%%%%%%%%%%%%%%%%%%%%%%%%%%%%%%%%%%%%%%%%%%%%%%%%%%
\subsection{Global picture}\label{subsec:global}%

\begin{figure}[t]
	\includegraphics[width=\textwidth]{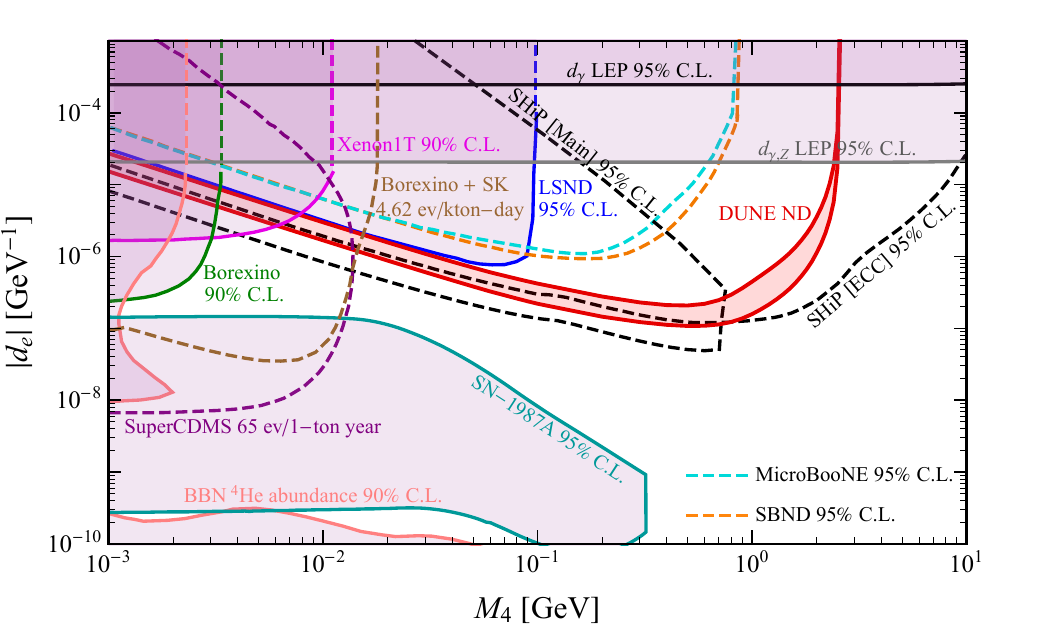}
	\caption{Comparison of the DUNE-ND sensitivity to $|d_e|$ obtained
	in this work (red solid) with current constraints (solid curves,
	shaded regions), and sensitivities of future projects or estimated
	exclusions (dashed curves). The band represents the region with 2
	-- 20 events/year, corresponding to 95\%~C.L.\ sensitivity over 5
	years with 25 -- 2500 background events. Limits and sensitivities
	are from LSND, MicroBooNE, SBND, SHiP, LEP, SN 1987A
	\cite{Magill:2018jla}, solar neutrinos \cite{Plestid:2020vqf,
	Brdar:2020quo}, Xenon1T, BBN \(^{4}\mathrm{He}\) abundance
	\cite{Brdar:2020quo}, SuperCDMS \cite{Shoemaker:2018vii}.}
	\label{fig:global-pic-e}
\end{figure}

\begin{figure}[t]
	\includegraphics[width=\textwidth]{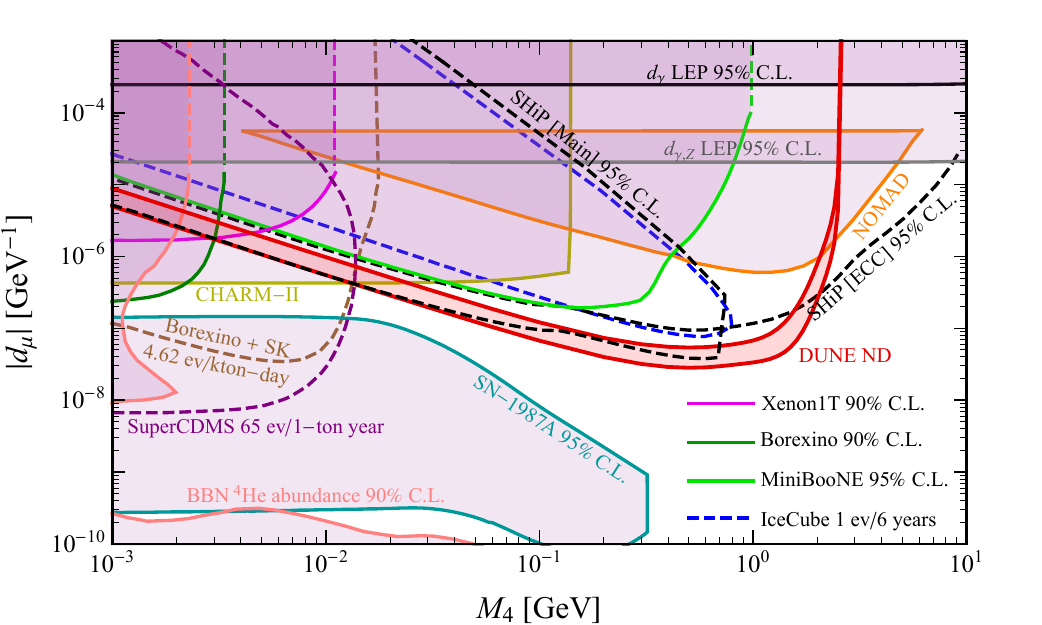}
	\caption{Comparison of the DUNE-ND sensitivity to $|d_\mu|$ obtained
	in this work (red solid) with current constraints (solid curves,
	shaded regions), and sensitivities of future projects or estimated
	exclusions (dashed curves). The band represents the region with 2
	-- 20 events/year, corresponding to 95\%~C.L.\ sensitivity over 5
	years with 25 -- 2500 background events. Limits and sensitivities
	are from CHARM-II \cite{Coloma:2017ppo,Geiregat:1989sz}, NOMAD
	\cite{Magill:2018jla,Gninenko:1998nn}, Icecube
	\cite{Coloma:2017ppo}, solar neutrinos
	\cite{Brdar:2020quo,Plestid:2020vqf}, MiniBooNE, SHiP, LEP, SN 1987A
	\cite{Magill:2018jla}, Xenon1T, BBN \(^4{\rm He}\) abundance
	\cite{Brdar:2020quo}, SuperCDMS \cite{Shoemaker:2018vii}.}
	\label{fig:global-pic-mu}
\end{figure}

\begin{figure}[t]
	\includegraphics[width=\textwidth]{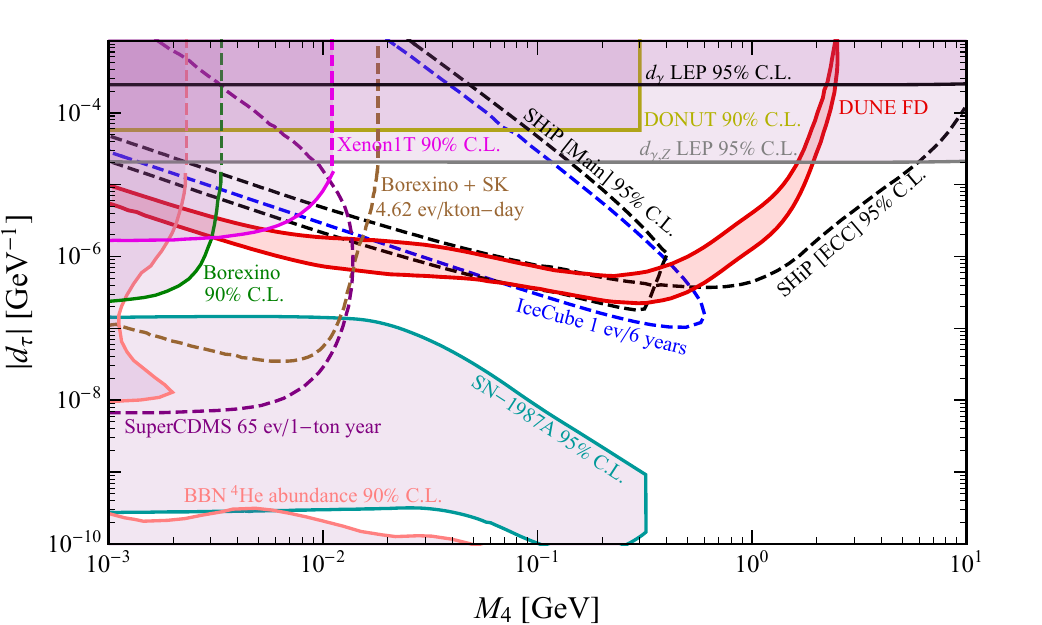}
	\caption{Comparison of the DUNE-FD sensitivity to $|d_\tau|$
	obtained in this work (red solid) with current constraints (solid
	curves, shaded regions), and sensitivities of future projects or
	estimated exclusions (dashed curves). The band represents the region
	with 2 -- 20 events/year, corresponding to 95\%~C.L.\ sensitivity
	over 5 years with 25 -- 2500 background events. Limits and
	sensitivities are from DONUT \cite{Schwienhorst:2001sj}, Icecube
	\cite{Coloma:2017ppo}, solar neutrinos \cite{Brdar:2020quo,
	Plestid:2020vqf}, SHiP, LEP, SN 1987A \cite{Magill:2018jla},
	Xenon1T, BBN \(^{4}\mathrm{He}\) abundance \cite{Brdar:2020quo},
	SuperCDMS \cite{Shoemaker:2018vii}.}
	\label{fig:global-pic-tau}
\end{figure}

Let us now summarise various constraints and sensitivities on the HNL
dipole portal and set our results for the DUNE FD and ND in the global
context. In
\cref{fig:global-pic-e,fig:global-pic-mu,fig:global-pic-tau} we show
the landscape of current constraints and projected sensitivities of
\(d_\alpha\) versus \(M_4\).\footnote{The magnetic moments of SM
  neutrinos may also lead to observable corrections in all kinds of
  neutrino experiments and some astronomical and cosmological
  processes.  The relevant discussions and stringent limits can be
  found in \cite{Giunti:2014ixa,Giunti:2015gga}.} Constraints from
previous experiments are shaded with solid boundaries and
sensitivities based on future experiments and estimated exclusions
(for which there is no rigorous background/selection-efficiency
analysis) are shown with dashed lines except for the DUNE ND and FD
(being highlighted in solid lines), the results of this work. In order
to illustrate the impact of background events, our results are
presented as a band, showing the region with 2 to 20
events/year. Using signal/$\sqrt{\text{background}} \approx 2$ this
would correspond to a 95\%~C.L.\ sensitivity over 5 years for the
range of 25 to 2500 background events. As discussed above, background
and efficiency considerations will be rather different for the
different classes of events (outside/inside coherent/inside
incoherent), a subtlety which we ignore here. A detailed study along
these lines is beyond the scope of this work; nevertheless the bands
shown in the figures of this section gives a rough indication of the
potential impact of background events or selection efficiencies.

We summarise below the various constraints and projected sensitivities in
\cref{fig:global-pic-e,fig:global-pic-mu,fig:global-pic-tau}:
\begin{itemize}
	\item We show the dominant constraints from LSND and MiniBooNE of
		$d_e$ and $d_\mu$, respectively \cite{Magill:2018jla}. The
		projected sensitivities to $d_e$ from SBND and MicroBooNE are also
		shown; for \(d_\mu\) they are similar to MiniBooNE, and are not
		shown, see \cite{Magill:2018jla}.
	\item By considering induced elastic scattering of $\nu_{\mu}$
		($\overline\nu_{\mu}$) on electrons, CHARM-II can constrain
		$d_\mu$ (dark yellow) \cite{Coloma:2017ppo, Geiregat:1989sz}. In
		a similar fashion, DONUT (an accelerator experiment dedicated to
		investigate tau-neutrino interactions) gave an upper limit on
		\(d_\tau\) \cite{Schwienhorst:2001sj}, which applies for
		\(M_4 \lesssim \SI{0.3}{GeV}\) (due to kinematics).
	\item At the SPS at CERN, the past NOMAD experiment searched for
		the single-photon signal from the up-scattering of
		neutrinos into sterile neutrinos (constraint from
		\cite{Magill:2018jla,Gninenko:1998nn} in orange in
		\cref{fig:global-pic-mu}). The proposed SHiP
		detector \cite{Alekhin:2015byh} (also at the SPS) consists of an
		emulsion cloud chamber (ECC) near detector and a main detector.
		Sensitivities from \cite{Magill:2018jla} are shown as black dashed
		curves. For electron and tau flavours a background of 100 events is
		assumed, while for the muon flavour a background level of 1000 events is
		assumed.
	\item We also show bounds calculated in \cite{Magill:2018jla} from
		LEP; however these depend on the UV-completion of the model, as
		above the electroweak scale the dipole operator must couple to the
		fields before electroweak symmetry breaking, which allows on-shell
		\(Z\) or \(W\) production. The solid black curves ignore on-shell
		\(W\) and \(Z\) production, while the solid grey include on-shell
		\(Z\)s. (See also Table II and Fig.\ 9 of \cite{Magill:2018jla}.)
	\item Constraints from modifications of the solar-neutrino electronic recoil spectrum
		at Borexino due to transition magnetic moments \cite{Grimus:2002vb,
		Borexino:2017fbd} were calculated for the HNL portal in
		\cite{Brdar:2020quo} and are shown in dark-green in
		\cref{fig:global-pic-e,fig:global-pic-mu,fig:global-pic-tau}. An
		analysis of solar-neutrino nuclear recoils from the Xenon1T
		dark-matter experiment leads to the constraints on $d_\alpha$
		shown in magenta \cite{Brdar:2020quo}. Constraints from the same
		type of signal at the future SuperCDMS dark matter detector are
		shown in purple assuming a 1-ton year exposure time and
		\(3\sigma\)-significance (65 events)
		\cite{Shoemaker:2018vii}. Due to oscillations, there is an
		\(\mathcal O(1)\) correction to the \(d_e\) constraints.
	\item A recent analysis \cite{Plestid:2020vqf} considers a situation,
		similar to ours, of up-scattering of solar neutrinos in the Earth and decay in the
		detector; it combines data from both Borexino
		\cite{Agostini:2018uly} and Super-Kamiokande (SK)
		\cite{Abe:2016nxk} to derive the constraints of \(d_{\alpha}\)
		(brown, dashed); however this analysis does not consider the geometric
		suppression rigorously for outside events.
	\item From the observed SN 1987A neutrino burst limits on \(d_
		{\alpha}\) can be set \cite{Magill:2018jla}. The areas enclosed by
		the cyan curves in
		\cref{fig:global-pic-e,fig:global-pic-mu,fig:global-pic-tau} are
		disfavoured by SN 1987A. Below the curve, the induced cooling
		effect is too weak, and above the interaction becomes strong enough
		so that steriles cannot escape the collapsing core. Finally, if
		the sterile is too heavy, the gravitational pull will also prevent
		it from leaving the supernova, leading to the vertical cut-off of
		the exclusion curve.
	\item In \cite{Coloma:2017ppo}, bounds from double-bang events (a
		signal with two visibly separate cascades) at IceCube from the
		atmospheric neutrino flux were calculated.
		These are denoted by the blue dashed lines in
		\cref{fig:global-pic-mu,fig:global-pic-tau}, namely one event
		during a data-taking period of six years.
	\item Bounds from cosmology and Big-Bang Nucleosynthesis (BBN) are
		shown in pink. The dipole interaction alters the expansion and
		cooling rates of the universe, leading to a corrected
		neutron-to-proton ratio and baryon-to-photon ratio
		\cite{Brdar:2020quo}. The final \(^4{\rm He}\) abundance depends
		on \(M_4\) and \(d_\alpha\). For the observed primordial-\(^4{\rm
		He}\) fraction \(Y_p=0.245\pm0.006 ~(2\sigma)\), the corresponding
		constraints of \(d_\alpha\) are illustrated in
		\cref{fig:global-pic-e,fig:global-pic-mu,fig:global-pic-tau}.
\end{itemize}

%%%%%%%%%%%%%%%%%%%%%%%%%%%%%%%%%%%%%%%%%%%%%%%%%%%%%%%%%%%%%%%%%%%%%%%
\section{DUNE-FD sensitivity to active-sterile mixing via \texorpdfstring{$U_{\tau 4}$}{}}
\label{sec:mixing}%

So far we assumed that the mixing of the sterile neutrino with active
flavours is negligible compared to the dipole interaction. In this
section we briefly comment on the opposite case when the heavy
neutrino mixes with strength $|U_{\alpha 4}|^2$ with active neutrinos
and the dipole interaction is negligible. There are strong bounds on
mixing with electron and muon flavour, as well as excellent prospects
for upcoming projects, see for example \cite{Atre:2009rg,
Ballett:2016opr,SHiP:2018xqw,Bondarenko:2018ptm,Berryman:2019dme,
Coloma:2020lgy,Breitbach:2021gvv}. Therefore, we focus on mixing with
the tau flavour $|U_{\tau 4}|^2$, which is more difficult to probe.

We explore a similar phenomenology as in the case of the dipole: we
use the $\nu_\tau$ flux generated by $\nu_\mu\to\nu_\tau$ oscillations
in the DUNE experiment.\footnote{For similar considerations in the
context of solar and atmospheric neutrinos see \cite{Plestid:2020ssy,
Coloma:2019htx}.} These neutrinos can interact via NC interactions in
the Earth or inside the far detector and up-scatter to $\nu_4$.
Instead of the massless photon mediator, up-scattering is mediated by
\(Z^0\)-exchange. As previously mentioned, since the mediator is
massive, the total cross section depends linearly on the target mass
(rather than logarithmically as in the dipole case) and therefore
up-scattering on electrons can be neglected.  Subsequently the heavy
neutrino can decay and leave an observable signal in the detector. The
main decay processes for \(M_4<\SI1{GeV}\) and mixing only with the
tau flavour are \(\nu_4\to\nu_\tau\nu_l\bar \nu_l\), \(\nu_4\to
\nu_\tau e^+e^-\) and \(\nu_4\to\nu_\tau\pi^0\), the latter two
providing a visible signal. There is considerable confusion in the
literature about the decay widths of a sterile neutrino that mixes
with SM neutrinos with disagreeing results, compare \emph{e.g.\
}\cite{Orloff:2002de,Atre:2009rg,Ballett:2016opr,Bondarenko:2018ptm,
Coloma:2020lgy}. In our work we use the formulae presented recently in
\cite{Coloma:2020lgy} where a discussion on various previous results
can also be found.

\begin{figure}[t]
	\includegraphics{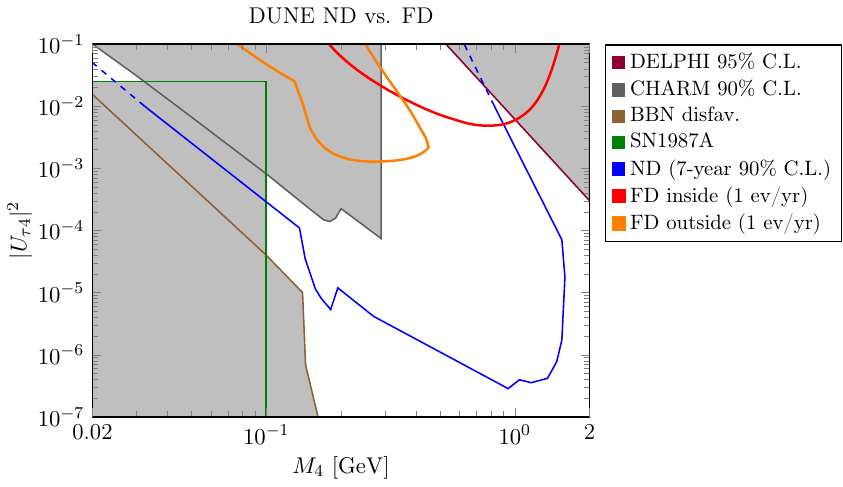}
	\caption{Comparison of the one-event/yr curves (100\% efficiency)
	for inside and outside events at the DUNE FD with the 7-year
	90\%~C.L.\ sensitivity at the DUNE ND assuming a 20\% selection
	efficiency and no background from \cite{Coloma:2020lgy}. The ND
	curve has been extrapolated to higher mixings (dotted line). We
	also show the existing limits from DELPHI
	\cite{Abreu:1996pa,Atre:2009rg} and CHARM~\cite{Orloff:2002de} as
	well as BBN and SN 1987A constraints from \cite{Dolgov:2000jw}.}
	\label{fig:DUNE-NC}
\end{figure}%

In \cref{fig:DUNE-NC} we show the curves in the plane of
heavy-neutrino mass $M_4$ and mixing parameter $|U_{\tau 4}|^2$
corresponding to 1 signal event/year in the DUNE FD separately for
inside and outside events. We see that the sensitivity covers a small
region at relatively large mixing between the current exclusion
limits from DELPHI \cite{Abreu:1996pa,Atre:2009rg} and CHARM \cite{
Orloff:2002de}. Let us note that a disagreement in the decay rate for
$\nu_4\to\pi^0\nu_\tau$ used in \cite{Orloff:2002de} compared to
recent papers (\emph{e.g.\ }\cite{Bondarenko:2018ptm,Coloma:2020lgy})
may potentially effect the CHARM bound shown in the figure.

Also shown in the figure is the sensitivity of the DUNE near detector
from \cite{Coloma:2020lgy} (90\%~C.L.\ sensitivity over 7 years of
running, or \SI{7.7e21}{POT}). This signal comes from the (small ---
though non-zero) sterile-neutrino flux produced by meson decays in the
target to probe heavy-neutrino mixing with the tau flavour. The
authors assumed a 20\% selection efficiency with no background
(corresponding to 2.44 events over 7 years for 90\%~C.L.\
sensitivity).\footnote{Recently a similar study has been performed in
\cite{Breitbach:2021gvv} including more realistic detector simulations
and background considerations. In \cite{Berryman:2019dme} the
sensitivity to HNL mixing (including the tau flavour) of a so-called
``multi-purpose'' near detector at DUNE has been considered.}
Assuming the same selection efficiency, our one-event curve
corresponds to \(7\times20\%=1.4\) events over 7 years. As obvious
from this figure, for the case of mixing the FD cannot compete with
the ND. The difference with the dipole case is that the event rate in
the FD relative to the ND is suppressed by the NC cross section
involved in the up-scattering. A summary of the sensitivity to this
scenario of various other planned projects (including SHiP) can be
found e.g.\ in \cite{Breitbach:2021gvv}.

BBN constraints from \cite{Dolgov:2000jw} are also shown in \cref{fig:DUNE-NC},
corresponding to \(\Delta N_\text{eff}= 0.2\). For masses \(M_4\) larger than
the neutral-pion mass, the bound weakens due to the large visible decay width.
From ibid.\ bounds are also derived from the SN 1987A observation based on the
thesis that any emission into exotic dark sectors would reduce the duration of
the neutrino burst, which is \(\sim\mathcal O(\SI1s)\) (see also
\cite{Raffelt:1999tx, Dolgov:2000pj}). The bound is
\(|U_{\tau4}|^2\lesssim\num{7.5e-9}\) for \(M_4\lesssim\SI{100}{MeV}\); but if
the sterile neutrino's decay length is too short, it never leaves the collapsing star's
core and doesn't contribute to the energy loss. The authors estimate this to
occur when \(|U_{\tau4}|^2\gtrsim\num{2.5e-2}\), where we cut off the bound.

%%%%%%%%%%%%%%%%%%%%%%%%%%%%%%%%%%%%%%%%%%%%%%%%%%%%%%%%%%%%%%%%%%%%%%%
\section{Conclusion}\label{sec:conclusion}%

We have calculated the sensitivity of the DUNE experiment to the
dipole portal with a heavy neutral lepton. Given the neutrino beam
energy, DUNE can probe HNLs in the mass range from a few MeV to a few
GeV, with a peak sensitivity for masses $M_4 \simeq 300$ to 500~MeV.
We focus here on HNL production via the up-scattering of active
neutrinos due to photon-mediated interactions with the matter
surrounding the detector or inside the detector (for sufficiently
short-lived sterile neutrinos). The signal consists of single-photon
events either with or without a displaced NC-type interaction
(depending on whether the up-scattering happens inside or outside the
detector and whether the scattering happens coherently on the full
nucleus).

We have shown that the far detector provides a competitive limit for
the transition magnetic moment between the tau neutrino and the HNL,
thanks to the sizeable tau neutrino flux generated by
$\nu_\mu\to\nu_\tau$ oscillations. DUNE will be able to probe the
tau-neutrino dipole portal down to $d_\tau$ of order a few~$\times
10^{-7}\,{\rm GeV}^{-1}$. Electron- and muon-neutrino dipole portals
are more efficiently probed at the near detector, thanks to the large
flux of these flavours. Restricting our analysis to
up-scattering-induced events, we find sensitivities for $d_e$ down to
$10^{-7}\,{\rm GeV}^{-1}$ and a few~$\times 10^{-8}\,{\rm GeV}^{-1}$
for $d_\mu$. Our results for DUNE are compared in \cref{subsec:global}
to various present constraints and upcoming sensitivities available in
the literature.

Motivated by these results, one may ask the question, whether the
oscillation-induced flux of tau neutrinos at the far detector can also
be used to test HNL \emph{mixing} with $\nu_\tau$. While we find
indeed that the far detector can test a currently unconstrained
region, it turns out that in this case the signal in the near detector
from the prompt HNL flux dominates. The reason is the additional
suppression of the far detector flux due to the NC cross section
required for the $\nu_\tau \to$~HNL up-scattering.

As an outlook, let us mention some topics for future work. First, we
concentrate here on estimating the size of the signal. A reliable
sensitivity calculation needs to take into account a detailed
background analysis, and making use of event discrimination abilities
in the liquid argon detector. Second, we have focused here on the HNL
production via the up-scattering of active neutrinos mediated by the
magnetic-moment interaction. In principle, the same interaction will
also lead to HNL production in the beam target from meson decays via
virtual-neutrino or -photon exchange. These HNLs may contribute to the
signal in the near detector. Predicting such a signal requires a
detailed simulation of the meson production and decay in the beam
target and decay pipe, which is beyond the scope of the present work.

To conclude, a transition magnetic moment $d_\alpha$ between active
and sterile neutrinos provides an attractive portal to search for
physics beyond the Standard Model. We have shown that the DUNE
experiment in its standard configuration has promising potential to
test all three flavours, $d_{e,\mu,\tau}$, when considering both the
near and far detectors. Our results warrant more detailed studies in
terms of background estimates and signal discrimination methods.%

\acknowledgments %%%%%%%%%%%%%%%%%%%%%%%%%%%%%%%%%%%%%%%%%%%%%%%%%%%%%%%%%%%%%

We thank Mahmoud Al-Awashra for collaboration at the initial stages of
this project and Ryan Plestid for useful discussions. Albert Zhou
thanks the Doctoral School ``Karlsruhe School of Elementary and
Astroparticle Physics: Science and Technology (KSETA)'' for financial
support through the GSSP program of the German Academic Exchange
Service (DAAD). This project has received support from the European
Union’s Horizon 2020 research and innovation programme under the Marie
Sklodowska-Curie grant agreement No 860881-HIDDeN. Jing-yu Zhu is
supported partly by the China and Germany Postdoctoral Exchange
Program from the Office of China Postdoctoral Council and the
Helmholtz Centre (Grant No.\ 2020031) and by the National Natural
Science Foundation of China (Grant No.\ 11835005 and 11947227).%

\paragraph{Postscript}
After the completion of this work the preprint \cite{Atkinson:2021rnp}
appeared, with partially similar considerations. Those authors
consider what they call \emph{double-bang} events, when both the
hadronic and single-photon decay signal are visible. These would be a
subclass of events considered here. Comparing with their DUNE
analysis, where our results overlap we find good agreement.%

%%%%%%%%%%%%%%%%%%%%%%%%%%%%%%%%%%%%%%%%%%%%%%%%%%%%%%%%%%%%%%%%%%%%%%%%%%%%%%%%%%%%%%%%%%%%
\appendix%
%%%%%%%%%%%%%%%%%%%%%%%%%%%%%%%%%%%%%%%%%%%%%%%%%%%%%%%%%%%%%%%%%%%%%%%%%%%%%%%%%%%%%%%%%%%%
\section{Cross sections} \label{app:xsec}%

We summarise here the relevant scattering cross sections \cite{Magill:2018jla, Brdar:2020quo}:
\begin{gather}
	\begin{split} \frac{\d\sigma(\nu_\tau e^-\to\nu_4e^-)}{\d Q^2}=
		\alpha_\text{QED}\left(\frac{|d_\alpha|}{\si{GeV}^{-1}}\right)^2
		\left[\frac2{Q^2}-\inv{m_{e}E_\nu}+\frac{M_4^2}{2m_{e}Q^2E_\nu^2}
		\times\right. \\ \left.\left(E_r-m_{e}-2E_\nu+M_4^2
		\frac{E_r-m_{e}}{Q^2}\right)\right]
		\times\SI{3.894e-28}{cm^2/nucleon} \\ \equiv X(m_{e});
	\end{split}
	\\
	\frac{\d\sigma_\text{nucleon}}{\d Q^2}=\frac{\d\sigma(\nu_\tau
	p\to\nu_4p)} {\d Q^2}+\frac{\d\sigma(\nu_\tau n\to\nu_4n)}{\d Q^2};
	\\
	\begin{split}
		\frac{\d\sigma(\nu_\tau p/n\to\nu_4p/n)}{\d Q^2}=F_{1,p/n}^2X\left
		(m_{p/n}\right)+\alpha_\text{QED}\left(\frac{|d_\alpha|}{\si{GeV}^
		{-1}}\right)^2\mu_{\mathrm N}^2\times \\ \left(\frac{F_2^{p/n}}{E_
		\nu}\right)^2\bigg[2(2E_\nu-E_r)^2-2Q^2+\frac{M_4^2}{m_{p/n}}\left(
		E_r-4E_\nu+M_4^2/E_r\right)\bigg]\\\times\SI{3.894e-28}{cm^2/nucleon};
	\end{split}
	\\
	\frac{\d\sigma(\nu_\tau N\to\nu_4N)}{\d Q^2}=\frac{2Z^2}AF_\text{nucleus}^2X(M_N)
\end{gather}
where \(E_r\equiv\frac{Q^2}{2M_T}\) is the recoil energy and \(\mu_{\mathrm N}\equiv \frac e
{2m_p}\approx0.16/\si{GeV}\) is the nuclear magneton. The numerical constant arises from the
conversion from \si{GeV^{-2}} to \si{cm^2}. Note that we normalise cross sections always per
\emph{nucleon}, not nucleus. This implies that the incoherent cross sections are a factor 2
less than na\"ively expected, since on average there are twice as many nucleons than protons,
neutrons and electrons. We also note that compared to other publications, we use \(d\equiv 2
\mu_\text{tr}\). The relevant form-factors are
\begin{equation}
	\begin{aligned}
		F_1^p &= \left(1 + \frac\eta{1+\eta}a_p\right)G_D, &F_1^n &= \frac\eta{1+\eta}a_nG_D \\
		F_2^{p/n} &= \frac{a_{p/n}}{1+\eta}G_D, &G_D &= \left(1+\frac{Q^2}{\SI{0.71}{GeV^2}}
		\right)^{-2} \\ a_p &= \mu_p-1 &a_n &= \mu_n,\quad\eta\equiv\frac{Q^2}{4M_T^2}.
	\end{aligned}
\end{equation}
for the nucleon, where \(a_{p/n}\) the anomalous magnetic moments of the proton/neutron:
\(1.793\) and \(-1.913\), respectively; and finally
\begin{equation}
	\begin{aligned}
		F_\text{nucleus}=\frac{3j_1(QR_\text{nucleus})}{QR_\text{nucleus}}\exp\left[-\frac{\left(
		Qs\right)^2}{2}\right] \\
		R_\text{nucleus}=\sqrt{\left(1.23\sqrt[3]A-0.6\right)^2+\frac73(\pi a)^2-5s^2}\si{fm}\\
		s=\SI{0.9}{fm}\quad a=\SI{0.52}{fm}
	\end{aligned}
\end{equation}
for the nucleus. \cite[§4]{Lewin:1995rx}%

\paragraph{Kinematic relations}
For the inside events, the spectrum is evaluated as a function of \(E_\nu\).
In order to convert from \(\d/\d Q^2\) to \(\d/\d\cos\theta_s\) we use kinematics to obtain
\begin{equation} \label{eq:E_nu-E_4-quad-eq}
	(E_\nu+M_T)E_4 - \left(M_TE_\nu+M_4^2/2\right) = E_\nu p_4\cos\theta_s.
\end{equation}
Define the following relations
\begin{equation}
	\begin{aligned}
		A_\pm&=M_T+E_\nu(1\pm\cos\theta_s), & B&=M_TE_\nu + M_4^2/2, \\
		\Delta&=1-(M_4/B)^2A_+A_-, & C&=\sqrt\Delta, \\
		E_4^0 &= B/A_-,&&
	\end{aligned}
\end{equation}
where $E_4^0$ corresponds to the heavy-neutrino energy in the limit
\(p_4=E_4\). The kinematic constraints are \(M_4<E_4<E_\nu\) and \(\Delta\geq0\).
This can be used to derive
\begin{equation}
	Q^2 = 2M_T\left[\frac{E_\nu^2(1-\cos\theta_s) - M_4^2/2}{A_-} +
	\frac{B(1-C)}{A_+A_-}E_\nu\cos\theta_s\right]
\end{equation}
and
\begin{equation} \label{eq:dQsq_dcos_th_s}
	\frac{\d Q^2}{\d\cos\theta_s}=-2M_T\frac{\d E_4}{\d\cos\theta_s}=\frac{2M_TE_\nu}{A_+}\left[
	\frac{2E_4E_\nu\cos\theta_s}{A_-}+E_4^0C+\frac{\left(M_4E_\nu\cos\theta_s\right)^2}{A_-BC}
	\right].
\end{equation}
When \(\Delta\) is very close to one, we expand in a Taylor series;
the same goes for \(p_N\) and other quantities with potentially small numbers.

\bigskip

For the outside events, the spectrum is evaluated as a function of \(E_4\). From \cref
{eq:E_nu-E_4-quad-eq}, one derives
\begin{equation}
	E_\nu=\frac{M_TE_4-M_4^2/2}{M_T-E_4+p_4\cos\theta_s}.
\end{equation}
Then (from \cref{eq:dQsq_dcos_th_s}),
\begin{equation}
	\frac{\d\sigma}{\d\cos\theta_s}\frac{\d E_\nu}{\d E_4}=-2M_T\frac{\d\sigma}{\d Q^2}\frac{\d
	E_\nu}{\d\cos\theta_s}=\frac{\d\sigma}{\d Q^2}\times\frac{2M_TE_\nu^2p_4}{M_TE_4-M_4^2/2}.
\end{equation}%

%%%%%%%%%%%%%%%%%%%%%%%%%%%%%%%%%%%%%%%%%%%%%%%%%%%%%%%%%%%%%%%%%%%%%%%%%%%%%%%%%%%%%%%%%%%
\section{Inside-event rate integral} \label{app:ins-ev-intgrl}%

We provide here some details on the calculation of the integral to
evaluate the inside-event rate. We focus on the far detector signal
due to $d_\tau$. Modifications for the near detector signal are
mentioned in \cref{subsec:near-det}.%

\paragraph{Neutrino flux}
For inside events we can restrict to the on-axis flux, which we obtained from
\cite{DUNE-off-axis-flux} in GLoBES format \cite{Huber:2007ji}. (These files are no longer
available as of the writing of this article; please see \cite[Fig.\ 5.4]{Abi:2020wmh}.) The
flux is provided for the near detector, therefore a geometric suppression factor
\(L_\text{ND}^2/|\vec x_p|^2\) was applied.%

\paragraph{Oscillation probability} Matter effects change the \(\nu_\mu\to\nu_\tau\)
oscillation probability only at the \(\mathcal O(1\%)\)-level. Therefore, we use the
effective-two-flavour vacuum probability
\begin{align} \label{eq:osc-prob}
	P_\text{osc}=0.943\cdot\sin^2\left(\frac{\Delta m^2|\vec x_p|}{4E_\nu}\right),&&\text{where}
	&& \Delta m^2=\SI{2.523e-3}{eV^2}.
\end{align}%

% \paragraph{Target density.} The DUNE far detector uses liquid argon
% (LAr), which has a density of \SI{1.396}{g/ml} \cite{LAr-BNL}. On the
% other hand, DUNE says the fiducial mass is \SI{10} {kt} with
% dimensions \(\SI{12}m\times\SI{14}m\times\SI {58.2}m\),\cite[Table
% 1.1]{Abi:2020loh} which gives a mass density of \(\rho_d=\SI{e10}g/(
% \SI{12}m\times\SI{14}m\times\SI{58.2}m)=\SI{1.02}{g/ml}\). We will use
% this number. We assume the discrepancy exists due to non-LAr
% components of the detector, like electronics. We will also assume the
% target density is constant in the detector.%

\paragraph{Decay probability} The decay width in the sterile neutrino's rest-frame is
\begin{equation}
	\Gamma_0 = \frac{|d_\alpha|^2M_4^3}{4\pi}
\end{equation}
and the probability to decay inside the detector is
\begin{equation}
	P_\text{dec} = 1 - \exp\left(-\gamma\ell_d\right),
\end{equation}
where
$\gamma\equiv\Gamma_0 M_4 / p_4$ is the lab-frame decay rate and
\(\ell_d\) is the distance from the production point \(\vec x_p\) to the edge of the
detector in the direction \(\vec x_p - \vec x_d\).%

\paragraph{Reconstruction efficiency} The reconstruction efficiency as a function of photon
momentum \(\varepsilon(p_\gamma)\) was taken from \cite[Fig.\ 4.26]{Abi:2020evt}. We assume
that in the sterile neutrino's rest frame the distribution of photon momentum is isotropic.
Boosting this into the lab-frame gives \(p_\gamma(\theta_0; p_4, M_4)\), where \(\theta_0\) is
the photon momentum's rest-frame polar angle. The efficiency, then, is
\begin{equation}\label{eq:efficiency}
	\varepsilon(p_4, M_4) = \int_0^{2\pi}\varepsilon[p_\gamma(\theta_0; p_4, M_4)]\d\theta_0 \,.
\end{equation}
Let us note that for a Dirac sterile neutrino, the differential decay
width \(\d\Gamma_0/\d\cos\theta_d\) is not isotropic in general
\cite{Balantekin:2018azf,Balantekin:2018ukw}. The asymmetry depends on
the relative complex phase of electric and magnetic dipole moments
and, in particular, it vanishes if one of them is zero. We have
checked, however, that even in the presence of an asymmetry the effect
on the integral in \cref{eq:efficiency} is at the few percent level
and has therefore negligible impact on our results.

%% For expediency, an effective parameterisation was used:
%% \begin{equation}
%% 	\varepsilon(p_4) = \begin{cases}
%% 		M_4\in[10^{-3},0.1]\si{GeV}\quad p_4\in[0,10]\si{GeV} \\
%% 		0.684327 + \sqrt{p_4}(0.239729 + \num{9.35499e-3} p_4 - 0.0858453 M_4) \\
%% 		+ \sqrt{M_4}[-\num{8.68688e-4} + 0.252934 M_4 + 0.0121437 p_4 \\
%% 		- \sqrt{p_4}(0.0116359 + 0.0907845 M_4 + \num{2.39247e-3} p_4)] \\
%% 		+ M_4(0.0323379 + 0.076513 M_4 + 0.0223495 p_4) \\
%% 		- p_4(0.0747159 + \num{3.05956e-4} p_4) \\[-2ex]
%% 		\rule{0.7\textwidth}{0.4pt} \\[-1ex]
%% 		M_4\in[0.1,5]\si{GeV}\quad p_4\in[0,10]\si{GeV} \\
%% 		0.677126 + \sqrt{p_4}(0.187919 - \num{1.83246e-3} p_4 + 0.0184768 M_4) \\
%% 		+ \sqrt{M_4}[0.168718 - 0.0250098M_4 + 0.0345782 p_4 \\
%% 		- \sqrt{p_4}(0.166654 - \num{6.33635e-3} M_4 + \num{9.4691e-4} p_4)] \\
%% 		+ M_4(0.0449037 - \num{1.79961e-6} M_4 - \num{7.15528e-3}p_4) \\
%% 		- p_4(0.0272767 - \num{3.71682e-4}p_4).
%% 	\end{cases}
%% \end{equation}%

\paragraph{Approximations and geometry}
Since the detector dimensions are \(\mathcal O(\SI{10}m)\), whereas
the baseline is \(L_\text{FD} \approx\SI{1300}{km}\) we can make the
approximations $|\vec x_p| \approx L_\text{FD}$ and ignore the beam
opening angle on the scale of the detector size, as the maximum angle
is \(\theta_b=\mathcal O(\SI{10}{m})/L_\text{FD}\sim\mathcal
O(10^{-5}) \approx 0\).

%% Extra care has to be taken regarding \(P_\text{osc}\propto\sin^2\frac{\Delta m^2x}{4E_\nu}\).
%% The \(\nu_\mu\to\nu_\tau\) transition occurs with \(\Delta m^2_\text{atm.}\sim
%% \SI{2.5e-3}{eV^2}\); even for energies as low as \(E_\nu\sim\SI{0.01}{GeV}\) the frequency of
%% \(P_\text{osc}\) is \(\Delta m^2/(4E_\nu)\sim\mathcal O (\SI{e-4}{m^{-1}})\), so we can safely
%% ignore the detector dimensions in the oscillation probability.%

\begin{figure}[t]\centering
	\includegraphics{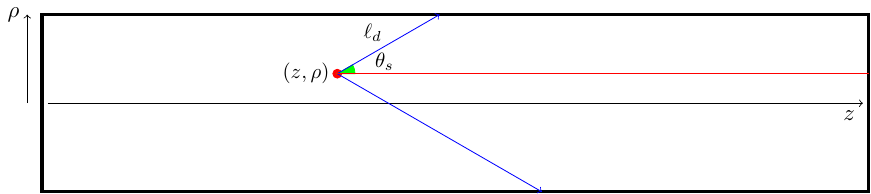}
%	\tikzsetnextfilename{decay-length}
%	\begin{tikzpicture}
%		\draw[ultra thick] (0,0)--(14,0)--(14, 3)--(0,3)--cycle;
%		\draw[->] (0.1,1.5)--(13.9,1.5) node [below left] {\(z\)};
%		\draw[->] (-0.25,1.5)--(-0.25,3) node [left] {\(\rho\)};
%		\filldraw[red] (5,2) circle [radius=0.075] node[black,left] {\small\((z, \rho)\)};
%		\filldraw[green] (5,2)--(5.3,2) arc [start angle=0, end angle=30, radius=0.3]--cycle;
%		\draw (5.8, 2.2) node [black] {\small\(\theta_s\)};
%		\draw[blue,->] (5,2)--(6.732,3);
%		\draw (5.6,2.7) node {\small\(\ell_d\)};
%		\draw[blue,->] (5,2)--(8.464,0);
%		\draw[red] (5,2)--(14,2);
%	\end{tikzpicture}
	\caption{An incoming neutrino scatters at \((z,\rho)\) in the
	cylindrical detector module. In blue are two possible paths for the
	sterile neutrinos, at constant scattering angle \(\theta_s\). In red
	is the trajectory of the neutrino, had it not scattered.}
	\label{fig:decay-length}
\end{figure}

We assume that the detector is cylindrical with radius \(r=\sqrt{\SI
{12} m\cdot\SI{14}m/\pi}\) and length \(L_d\approx\SI{58.2}m\). This
allows us to do the \(\varphi\) integral analytically. The geometry
determining the decay length is sketched in \cref{fig:decay-length}.
In the figure, the decay path ends at the sides of the detector. There could be the case when
both end at one of the ends of the detector. Then the decay-length is independent of the
scattering position, defined by \(\ell_d=(L_d-z)\sec\theta_s\) for the right end and \(\ell_d=
z\sec\theta_s\) for the left end. In this case our approximation \(\ell_d=\ell_d(\rho=0)\equiv
\ell_d^0\) is exact. We will not consider the case when the decay path ends partly on the side
and partly at the end, as this will only occur rarely in comparison to the other two cases,
and its effect will only be some kind of average between the two.

Let us consider the case, as drawn in the figure, when the decay path ends at the sides of the
detector. From the figure it is clear that \(\ell_d\sin\theta_s+\rho=r\). This is in the case
\(\varphi_s=0\). For non-zero \(\varphi_s\),
\begin{equation}
	(\ell_d\sin\theta_s\cos\varphi_s+\rho)^2 + \ell_d^2\sin^2\theta_s\sin^2\varphi_s=r^2.
\end{equation}
The solution to this quadratic equation (assuming \(\theta_s\in[0,\pi]\)) is
\begin{equation}
	\begin{aligned}
		\ell_d&=\left[\sqrt{r^2-\rho^2\sin^2\varphi_s}-\rho\cos\varphi_s\right]\cosec\theta_s \\
		&=\ell_d^0\left[\sqrt{1-x^2\sin^2\varphi_s}-x\cos\varphi_s\right],&& x\equiv\frac
		{\rho}r.
	\end{aligned}
\end{equation}
The penalty term (defined as the ratio of the exact decay probability
to the \(\rho=0\) approximation, \(P_\text{dec}(\ell_d)/
P_\text{decay}(\ell_d^0)\)) can then be written as
\begin{equation} \label{eq:Pi-penalty}
	\begin{aligned}
		\Pi\left(\gamma\ell_d^0\right)&=\left.\left(1-\frac{2\pi}{A_\text{det.}}\int_0^r\int_0^
		{2\pi}e^{-\gamma\ell_d}\frac{\d\varphi_s}{2\pi}\rho\d\rho\right)\right/\left(1-e^{-\gamma
		\ell_d^0}\right) \\
		&\begin{aligned}
			=\left.\left[1-\int_0^1\int_0^{2\pi}\exp\left(-\gamma\ell_d^0\left\{\sqrt{1-x^2\sin^2
			\varphi_s}-x\cos\varphi_s\right\}\right)\frac{\d\varphi_s}\pi x\d x\right]\right/ \\
			\hfill\left(1-e^{-\gamma\ell_d^0}\right).
		\end{aligned}
	\end{aligned}
\end{equation}
%% We use a parameterisation of \(\Pi(\,\cdot\,)\):
%% \begin{gather}
%% 	\Pi_P(x)=\left\{\begin{aligned}(I_0-I_1)\left|1-\frac x{x_0}\right|^{2.33}+I_1;&&
%% 		\begin{aligned}x\in[0,x_0]\\x_0=1.58\end{aligned}&&
%% 		\begin{aligned}I_0&=\Pi(0)\\I_1&=\Pi(x_0)\end{aligned} \\
%% 		\sum_{k=0}^5c_kx^k;&&x\in[x_0,10] \\
%% 		\begin{aligned}1-\left[C_0 + (C_0-C_1)e^{-\inv{(x-10)}}\right]\times\\
%% 		\frac{x+e^{-x}-1}{x^2}\end{aligned};&&
%% 		x\in[10,\infty];&&
%% 		\begin{aligned}
%% 			C_0 = 0.705096\\C_1 = 0.747
%% 		\end{aligned}
%% 	\end{aligned}\right. \\
%% 	\begin{aligned}
%% 		c_{5,4,3,2,1,0}=\num{-8.70947987e-06},\num{2.97597885e-04},\num{-3.86481570e-03}, \\
%% 		\num{2.22183672e-02},\num{-3.48615349e-02},\num{8.23104405e-01}.
%% 	\end{aligned}\nonumber
%% \end{gather}
%% The asymptotic form of \(\Pi_P(x\to\infty)\) is determined by evaluating \cref{eq:Pi-penalty}
%% with \(\ell_d=\ell_d^\text{min.}=r-\rho\), and using an arbitrary slowly-varying constant
%% (interpolating between \(C_0\) and \(C_1\)).

Finally we must calculate \(\ell_d^0\); the decay path ends at the sides of the detector if
\(l_d\cos\theta_s\leq L_d-z\) and \(\theta_s<\pi/2\), or \(l_d|\cos\theta_s|\leq z\) and
\(\theta_s\geq\pi/2\) (equivalently \(l_d\cos\theta_s>-z\)). Therefore,
\begin{equation}
	\ell_d^0=\begin{cases}\min\left[r\cosec\theta_s,(L_d-z)\sec\theta_s\right] &
		\text{if}\quad\cos\theta_s\geq0 \\
		\min\left(r\cosec\theta_s,z|\sec\theta_s|\right) &
		\text{if}\quad\cos\theta_s<0 .
	\end{cases}
\end{equation}

%%%%%%%%%%%%%%%%%%%%%%%%%%%%%%%%%%%%%%%%%%%%%%%%%%%%%%%%%%%%%%%%%%%%%%%%%%%%%%%
\section{Outside-event rate integral} \label{app:out-ev-intgrl}%

Here we provide some details for the outside-event rate calculations.
Again we focus on the far detector signal due to $d_\tau$;
modifications for the near detector signal are mentioned in
\cref{subsec:near-det}. Since evaluation of the cross section can be
expensive, all terms in the integrand which depend purely on \((E_\nu
,E_4)\) were pre-calculated and then stored in a bilinear
interpolator. This table had to be recalculated for each mass
\(M_4\). The dependence on the decay width, which is non-linear, was
also separated and recalculated for each \(d_\tau\). The CQUAD GSL
routine was used to undertake the integrals \cite{GSL}. The C++ source
code can be provided upon request to Albert Zhou.%

\paragraph{Off-axis flux} This case we take into account the different neutrino fluxes as a
function of the off-axis angle. The off-axis (GLoBES) fluxes were obtained from
\cite{DUNE-off-axis-flux,Abi:2020wmh}. We take \SI{62.72} {mrad} as the maximum beam angle.
Those files have units \si{GeV^{-1}m^{-2}POT^{-1}}. Bilinear interpolation is used to
interpolate between neutrino energies and beam angles.%

\paragraph{Evaluation of the \(\varphi_b\)-integral}%

\begin{figure}[t]
	\centering
	\includegraphics{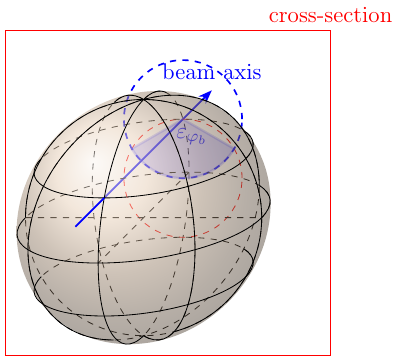}
	\includegraphics{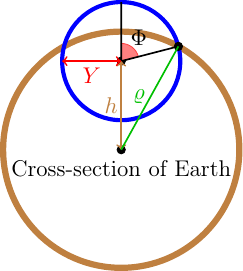}
	\caption{Example cross section with limits in \(\varphi_b\). The function \(\varepsilon_
	{\varphi_b}\) determines the relative contribution of the angles \(\varphi_b\) inside the
	Earth.}
	\label{fig:outside-3d-pic}
\end{figure}%

In \cref{fig:outside-3d-pic} (right), we see a cross section of the Earth, perpendicular to
the beam-axis. Due to cylindrical symmetry, the integration occurs in a plane perpendicular to
the radius of the Earth, with coordinates \((X, Y)=(r_p\cos\theta_b, r_p\sin\theta_b)\). This
plane is at a constant height \(h=\sqrt{R_\oplus^2-(L_\text{FD}/2)^2}\) above the Earth (with
the radius of the Earth \(R_\oplus=\SI{6371}{km}\). In the perpendicular cross section, the
rotation around \(\varphi_b\) produces a circle of radius \(Y\). The cross section shown is at
a distance \(\half{L_\text{FD}}-X\) away from the actual centre of the Earth (the \(X\)-axis
is perpendicular to the page). Thus, \(\varrho=\sqrt{R_\oplus^2-\left(\half{L_\text{FD}}-X
\right)^2}\) and \(\varepsilon_{\varphi_b}\equiv2(\pi-\Phi)=2\arccos\frac{Y^2+h^2-\varrho^2}
{2hY}\). In the limit \(Y=0\) however, we take \(\varepsilon_{\varphi_b}=2\pi\) for \(X\in[0,
L_\text{FD}]\) and 0 otherwise. Arc-cosine is only well-defined in the interval \([-1,1]\). If
its argument is \(>1\), this means there is no intersection between the blue and brown
circles, because the blue circle resides inside the brown: we set \(\varepsilon_{\varphi_b}=2
\pi\); for the case \(<-1\) the blue circle resides outside the brown one and \(\varepsilon_
{\varphi_b}=0\).%

\paragraph{Decay probability, scattering solid angle and detector geometry} Let \(\ell\) be
the distance from production \(\vec x_p\) to the detector \(\ell=\sqrt{L_\text{FD}^2+r_p^2-2L_
\text{FD}r_p\cos\theta_b}\); then the decay probability is
\begin{equation}
	P_\text{dec}=e^{-\gamma\ell}\left(1-e^{-\gamma L_{\text w}}\right),
\end{equation}
where we have taken the detector width \(L_{\text w}=\SI{14}m\) as the characteristic detector
dimension.

The scattering angle can be determined as
\begin{equation}
	\cos\theta_s=\frac{L_\text{FD}\cos\theta_b-r_p}\ell;
\end{equation}
this expression suffers from some numerical instability when the numerator has a larger
magnitude than the denominator, in which case the fraction on the RHS is set to its sign.

The scattering solid angle is
\begin{equation}
	\Delta\Omega_s=\sin\theta_s\Delta\theta_s\Delta\varphi_s=4\arctan\frac{L_d}{2\ell}\arctan
	\frac{L_{\text h}}{2\ell}\sin\theta_s,
\end{equation}
where \(L_d\), \(L_{\text h}\) are the length \SI{58.2}m and height \SI{12}m of the detector,
respectively. In the limit of large \(\ell\) (far away from the detector) and small \(\gamma
L_{\text w}\),
\begin{equation}
	P_\text{decay}\cdot\Delta\Omega_s=e^{-\gamma\ell}\sin\theta_s\gamma L_{\text w}L_dL_{\text
	h}/\ell^2\propto V_d/\ell^2
\end{equation}
we see the signal is proportional to the detector volume and its angular size \(\propto\ell^
{-2}\), as expected. Away from this limit, the geometric dependence inside the angular size
and decay probability will change how fast the signal decays; we neglect this affect.
Monte-Carlo studies with neutral-current up-scattering indicate that this effect results in a
\(\sim\) \numrange{20}{25}\% reduction of the spectrum (\emph{i.e.\ }a penalty of \numrange
{75}{80}\%). For the purpose of this study, we ignore this effect.

\paragraph{Oscillation phase}
For small \(E_4\), the spectrum \(\d N/\d E_4\) has fast wiggles for low \(E_4\). These arise
from the oscillation probability. Assuming \(E_4\sim E_\nu\), we set \(P_\text{osc}\) to
\(1/2\) when
\begin{equation}
	E_4<\frac{(\Delta m^2/4)L_\text{FD}}{4\pi},
\end{equation}
which is the fourth trough. The distance to the next trough is \(\frac{(\Delta m^2/4)L_
\text{FD}}{4\cdot5\cdot\pi}\sim\SI{6.6e-2}{GeV}\).%

\paragraph{Integration boundaries}
For masses \(M_4>\SI{e-2}{GeV}\) the integration boundaries are defined such that \(\ell<10/
\gamma\), determined by the exponential decay of the decay probability; for lower masses this
is not a good approximation. This results in
\begin{equation}
	\begin{aligned}
		\hat\ell\equiv\frac{\ell_\text{max}}{L_\text{FD}}&=\frac{100}{L_\text{FD}\gamma},\quad
		\Delta_G= \sqrt{\hat\ell^2 - \sin^2\theta_b}; \\
		\frac{r_\text{min.}}{L_\text{FD}}&=\begin{cases}
			0 & M_4\leq\SI{e-2}{GeV} \\
			\max\left(0, \cos\theta_b - \Delta_G\right) & M_4>\SI{e-2}{GeV},
		\end{cases} \\
		\frac{r_\text{max.}}{L_\text{FD}}&=\begin{cases}
			1.5 & M_4\leq\SI{e-2}{GeV} \\
			\cos\theta_b + \Delta_G & M_4>\SI{e-2}{GeV}.
		\end{cases}
	\end{aligned}
\end{equation}
Furthermore, the detector itself is removed for \(Y<\SI{15}m\), \(|X-L_\text{FD}|<L_d\).%

%%%%%%%%%%%%%%%%%%%%%%%%%%%%%%%%%%%%%%%%%%%%%%%%%%%%%%%%%%%%%%%%%
\section{Estimate of the near-detector signal} \label{app:dipole-ND}

In this appendix we provide a rough estimate of the signal induced in
the near detector by the $d_\tau$ transition moment. Such a signal
requires the presence of a prompt \(\nu_\tau\) flux at the
near detector. Such a flux is not available in the files provided by
DUNE at \cite{DUNE-off-axis-flux}; however it has been estimated
recently in \cite{Coloma:2020lgy,Breitbach:2021gvv} to study the
sensitivity of sterile neutrino mixing with the tau flavour. Here we
use their results to get a rough estimate for the dipole signal in the
ND.

In the limit of small mixing, the number of
mixing-induced sterile neutrino decays, $N_{\rm mix}$, can be estimated by
\begin{equation}\label{eq:NmixND}
	N_{\rm mix}\sim |U_{\tau4}|^2\Gamma_0\frac{M_4}{p_4}\Delta\ell_\text{det.} N_\tau \varepsilon \,,
\end{equation}
where $\Gamma_0$ is the rest-frame decay width and $N_\tau$ is the
number of $\nu_\tau$ passing through the detector during a given time
period. Using a typical momentum \(p_4\sim\SI{10}{GeV}\) (see \cite[Fig.\ 3]{Coloma:2020lgy})
and with the assumption that \(\nu_4\to\pi^0\nu_\tau\) dominates (see \cite[Fig.\
2]{Coloma:2020lgy}) we have
\(\Gamma_0=\frac{G_F^2M_4^3}{32\pi}f_\pi^2|U_{\tau4}|^2\left[1-\left(M_{\pi^0}/M_4\right)^2
\right]^2\). From Fig.~7 of \cite{Coloma:2020lgy} we can see that there are 2.44 events over 7
years at \(M_4=\SI{0.2}{GeV}\) and \(|U_{\tau4}|^2=10^{-5}\). Accounting for the 20\%
efficiency \(\varepsilon\) and the assumed 7 years of exposure we obtain from \cref{eq:NmixND}
\begin{equation}
	N_\tau\Delta\ell_\text{det.} \sim \SI{3.6e12}{m/yr}.
\end{equation}

Moving now to the case of dipole interaction, the
number of dipole decay events, $N_{\rm dip}$, can be estimated as
\begin{equation}
	N_{\rm dip} \sim \rho_N^\text{ND}\sigma_\text{tot}(N_\tau\Delta\ell_\text{det.}) \,,
\end{equation}
where we assume that all $\nu_4$ decay inside the detector. Using a
near-detector mass of 50 tonnes, a volume of
\(\SI3m\times\SI5m\times\SI7m\) \cite{AbedAbud:2021hpb} we obtain a
nucleon density of \SI{2.9e23}{cm^{-3}}. Taking for the dipole
induced cross section \(\sigma_\text{tot}\) from
\cite[eq.\ (3)]{Magill:2018jla} and \(Z=18\) for Argon (we also divide
by the number of nucleons \(A=40\)), we can estimate the signal for
the dipole portal as
\begin{equation}
	N_{\rm dip} \sim \num{1.8e-3}\left(\frac {d_\tau}{\SI{e-6}{GeV^{-1}}}\right)^2
	\text{yr}^{-1} \qquad (M_4 = \SI{0.2}{GeV})\,.
\end{equation}
Hence, this result suggests that the signal in the ND will only be
marginally relevant compared to the FD signal. Our estimate applies to inside events. Outside
events are expected to have the same order of magnitude.

%%%%%%%%%%%%%%%%%%%%%%%%%%%%%%%%%%%%%%%%%%%%%%%%%%%%%%%%%%%%%%%%%
\bibliographystyle{JHEP_improved}
\bibliography{./refs}
\end{document}